\documentstyle[12pt,epsf]{article}
\newcommand{\be}{\begin{equation}}
\newcommand{\ee}{\end{equation}}
\newcommand{\beqs}{\begin{eqnarray}}
\newcommand{\eeqs}{\end{eqnarray}}

\begin{document}
\pagestyle{plain}
\setcounter{page}{1}
\newcounter{bean}
\baselineskip16pt
 \begin{titlepage}
 \begin{flushright}
SU-ITP 97/51  \\
hep-th/9712072
\end{flushright}

\vspace{7 mm}

\begin{center}
{\huge Review of Matrix Theory }

\vspace{5mm}

{\huge }
\end{center}
\vspace{10 mm}
\begin{center}
{\large
D. Bigatti$^{\dagger}$ and L. Susskind } \\
\vspace{3mm}
Stanford University

\vspace{7mm}

{\large Abstract} \\
\end{center}

\vspace{3mm}
In this article we present a self contained
review of the principles of Matrix Theory
including the basics of light cone quantization,
the formulation of 11 dimensional M-Theory in
terms of supersymmetric quantum mechanics, the
origin of membranes and the rules of
compactification on 1,2 and 3 tori. We emphasize
the unusual origins of space time and gravitation
which are very different than in conventional
approaches to quantum gravity. Finally we discuss
application of Matrix Theory to the quantum
mechanics of Schwarzschild black holes. 

This work is based on lectures given by the second author at the 
Cargese ASI 1997 and at the Institute for Advanced Study in Princeton. 

\noindent

\vspace{7mm}
\begin{flushleft}
September 1997

\vspace{7cm} 
$\dagger$ On leave of absence from Universit\`a di Genova, Italy  

\end{flushleft}
\end{titlepage}
\newpage
\renewcommand{\baselinestretch}{1.1}


\renewcommand{\epsilon}{\varepsilon}
\def\fixit#1{}
\def\comment#1{}
\def\equno#1{(\ref{#1})}
\def\equnos#1{(#1)}
\def\sectno#1{section~\ref{#1}}
\def\figno#1{Fig.~(\ref{#1})}
\def\D#1#2{{\partial #1 \over \partial #2}}
\def\df#1#2{{\displaystyle{#1 \over #2}}}
\def\tf#1#2{{\textstyle{#1 \over #2}}}
\def\d{{\rm d}}
\def\e{{\rm e}}
\def\i{{\rm i}}
\def\Leff{L_{\rm eff}}

\def \td {\tilde }
\def \ci {\cite}
\def \sm {$\s$-model }

\def \o {\omega}
\def \inv {^{-1}}
\def \ov {\over }
\def \four{{\textstyle{1\over 4}}}
\def \fourth{{{1\over 4}}}
\def \ha {{1\ov 2}}
\def \QQ {{\cal Q}}

\def\sqr#1#2{{\vcenter{\vbox{\hrule height.#2pt
         \hbox{\vrule width.#2pt height#1pt \kern#1pt
            \vrule width.#2pt}
         \hrule height.#2pt}}}}
\def\square{\mathop{\mathchoice\sqr34\sqr34\sqr{2.1}3\sqr{
1.5}3}\nolimits}

\def\TL{\hfil$\displaystyle{##}$}
\def\TR{$\displaystyle{{}##}$\hfil}
\def\TC{\hfil$\displaystyle{##}$\hfil}
\def\TT{\hbox{##}}

\def\shortlistrefs{\footatend\bigskip\bigskip\bigskip%
Immediate\closeout\rfile\writestoppt
\baselineskip=14pt\centerline{{\bf
References}}\bigskip{\frenchspacing%
\parindent=20pt\escapechar=` Input
refs.tmp\vfill\eject}\nonfrenchspacing}

\def\eff{{\rm eff}}
\def\abs{{\rm abs}}
\def\hc{{\rm h.c.}}
\def\+{^\dagger}

\def\cl{{\rm cl}}

\def\M{\cal M}
\def\D#1#2{{\partial #1 \over \partial #2}}

\def\overleftrightarrow#1{\vbox{Ialign{##\crcr

\leftrightarrow\crcr\noalign{\kern-0pt\nointerlineskip}
     $\hfil\displaystyle{#1}\hfil$\crcr}}}

\def \t {\tau}
\def \td {\tilde }
\def \ci {\cite}
\def \sm {$\s$-model }

\def \o {\omega}
\def \inv {^{-1}}
\def \ov {\over }
\def \four{{\textstyle{1\over 4}}}
\def \fourth{{{1\over 4}}}
\def \ha {{1\ov 2}}
\def \QQ {{\cal Q}}

\def \lr { \lref}
\def\np {{  Nucl. Phys. }}
\def \pl {{  Phys. Lett. }}
\def \mpl {{ Mod. Phys. Lett. }}
\def \prl {{  Phys. Rev. Lett. }}
\def \pr  {{ Phys. Rev. }}
\def \ap  {{ Ann. Phys. }}
\def \cmp {{ Commun.Math.Phys. }}
\def \ijmp {{ Int. J. Mod. Phys. }}
\def \jmp {{ J. Math. Phys.}}
\def \cqg {{ Class. Quant. Grav. }}





\section*{Introduction  (Lecture zero)}
\par Matrix theory
 \cite{1}  is a nonperturbative theory of fundamental
processes which
evolved out of the older perturbative string theory. There are
two well-known
formulations of string theory, one covariant and one in the
so-called light
cone frame
 \cite{2}. Each has its advantages. In the covariant theory,
relativistic
invariance is manifest, a euclidean continuation exists and the
analytic
properties of the S matrix are apparent. This makes it relatively
easy to
derive properties like CPT and crossing symmetry. What is less
clear is that
the theory satisfies all the rules of conventional unitary quantum
mechanics.
In the light cone formulation
 \cite{3}, the theory is manifestly hamiltonian
quantum
mechanics, with a distinct non-relativistic flavor. There exists a
space of
states and operators which is completely gauge invariant and
there are no
ghosts or unphysical degreees of freedom. Among other things,
this makes it
easy to identify the spectrum of states. Matrix theory is also
formulated
in the light cone frame and enjoys the same advantages and
disadvantages as LCF string
theory. Unlike perturbative string theory, matrix theory is capable of
addressing many of the nonperturbative questions raised by
string theory and quantum gravity.
\par The need for a nonperturbative theory became very
apparent a few years ago
from two lines of research: black holes and duality. In the case of
black holes,
it was clear to a few of us that explaining the Bekenstein Hawking entropy
would require a microscopic description involving degrees of
freedom more fundamental than those that are available in
ordinary general relativity
 \cite{4}. Furthermore,
although string theory offered some insight, the weakly coupled theory could
never completely solve this problem since black holes are fundamentally non
perturbative objects. In addition, the so-called information paradox of Hawking
suggested that there is something very naive and wrong with our usual ideas
of locality that we inherited from quantum field theory. As we shall see,
matrix
theory does not seem to be a space-time quantum field theory even
approximately,
except in the low energy perturbative regime.
\par The need for a more powerful approach was made even clearer when it was
discovered that the various different string theories are really part of a vast
``web'' of solutions of a single 11-dimensional theory called M-theory
 \cite{5}. The web
is parametrized by the ``moduli'' describing the many different
compactifications. The perturbative string theories describe remote corners of
the moduli space, but most of the space is beyond their reach. Each string
corner has a spectrum of light weakly coupled states such as IIa, IIb, I or
heterotic strings. Perhaps the right way to say it is that all of the objects
are present in all of moduli space, but in each corner some particular set
becomes light and weakly coupled. However, perturbation theory in any of the
corners can not detect the other corners and the web that connects them.
\par It is evident that any fundamental degrees of freedom which can describe
all these stringy excitations as well as 11-dimensional gravitons, membranes,
5-branes, D-branes and black holes must be very special and unusual. An even
more surprising indication that the underlying degrees of freedom are unlike
anything previously encountered is their ability to grow new dimensions of
space. The simplest example is the compactification of the 11-dimensional
M-theory on a two torus
 \cite{6}. This obviously leads to a theory with 9 noncompact
dimensions. Now let the torus shrink to zero size. The surprise is that in this
limit a new 10th noncompact dimension appears. Nothing like this was ever
imagined in older ideas based on quantizing the classical gravitational field.
\par Thus it is very satisfying that a simple system of degrees of freedom
which does all of the above has been identified. In the following set of
lectures  we will describe the ``matrix theory'' of this system of protean
degrees of freedom and how it accomplishes the necessary feats.
The lectures are self contained and do not require an extensive knowledge of
string theory. However they far from exhaust the important and interesting things that have been learned in the last year. 

We would like to warn the reader that here and there an attempt has been made to keep track of numerical factors, but they should not be expected to be completely consistent. 


\section{Lecture 1 (The light cone frame)}
\par In this lecture the light cone method will be reviewed
 \cite{3} . The dimension of
space-time is $D$. The space-time dimensions are labeled
$X^\mu=(t, z, X^1, \cdots , X^{D-2} )$ \, .
One of the spatial coordinates $z$
has been singled out. It is called the longitudinal direction and the space
$(X^1, \cdots , X^{D-2} )$ is called the transverse plane. Now introduce
two light-like coordinated $X^\pm$ to replace $(t, z)$ :
\begin{eqnarray}  \displaystyle{
X^\pm =  \frac{t \pm z}{\sqrt{2}}
}  \end{eqnarray}
In the light cone frame $X^+$ is used as a time variable and the space $X^+=0$
is the surface of initial conditions. A quantum state in the Schr\"{o}dinger
picture describes data on a surface $X^+={\mathrm{constant}} $. A
Hamiltonian
is used to propagate the quantum state from one value of $X^+$ to another:
\begin{eqnarray}  \displaystyle{
|X_2^+ \rangle = e^{ i H (X_2^+ - X_1^+ ) }   |X_1^+ \rangle
}  \end{eqnarray}
\par Let us consider the form of $H$ for an arbitrary system with center
of mass energy $M$. The system need not be a single particle. Let us also
introduce
conjugate momenta $P_\pm = \frac{ (P_t \pm P_{z}) }
{ \sqrt{2}}$. The ``on
shell'' condition is
\begin{eqnarray}  \displaystyle{
P_\mu P^\mu = M^2
}  \end{eqnarray}
or
\begin{eqnarray}  \displaystyle{
2 P_+ P_- - P_i^2 = M^2
}  \end{eqnarray}
Now let us define $P_+$ to be the hamiltonian $H$. We find
\begin{eqnarray}  \label{(1)} \displaystyle{
H= \frac{P_i^2}{2 P_-} + \frac{M^2}{2 P_-}
}  \end{eqnarray}
Eq.~(\ref{(1)}) has a nonrelativistic look to it. Let's compare it with the
general expression for energy in a Galilean invariant theory in $D-2$ spatial
dimensions.
\par Let $\mu$ and $p$ be the total mass and momentum of the system and let the
Galilean invariant internal energy be called $U/2\mu$. Then the total energy is
\begin{eqnarray}  \label{2} \displaystyle{
H= \frac{p^2}{2\mu} + \frac{U}{2\mu}
}  \end{eqnarray}
Thus, if we identify the nonrelativistic mass of a system with its 
$P_-$ and $U$ with the
invariant $M^2$, the formulas are the same. Is this a coincidence or there is
an underlying reason for this? The answer is that there is a subgroup of the
Poincar\'e group, relevant for light cone physics, which is isomorphic to the
Galilean group.
\par Let us review the Galilean group. Classically it is generated by the
following:
\begin{eqnarray}  \nonumber \displaystyle{
\begin{array}{cl}
p \:\:\:\:\:\:\:\:\:\:  & \:\:\:\:\:\:\:\:\:\:
\mathrm{spatial\:translations} \\
h \:\:\:\:\:\:\:\:\:\: &  \:\:\:\:\:\:\:\:\:\:
\mathrm{time\:translations} \\
l_{ij}  \:\:\:\:\:\:\:\:\:\:   &    \:\:\:\:\:\:\:\:\:\:
{\mathrm{angular\:momentum\:in\:the\:}} ij  \: {\mathrm{ plane}} \\
\mu x^i_{C.M.}  \:\:\:\:\:\:\:\:\:\:  &  \:\:\:\:\:\:\:\:\:\:
\mathrm{Galilean \: boosts.}  \\
\end{array}
}  \end{eqnarray}
To see that $\mu x^i_{C.M.}$ generates boosts consider the action of
$exp(iV \cdot \mu x^i_{C.M.}) $ on $p$. From the fact that
$x_{C.M.}$ and $p$ are conjugate variables
\begin{eqnarray}  \displaystyle{
e^{[i V \cdot \mu x^i_{C.M.}]} \: p \: e^{-[i V \cdot \mu x^i_{C.M.}]}
= p + \mu V
}  \end{eqnarray}
which has the form of a boost with velocity $V$. 
\par One more generator belongs to the Galilean group. The commutation relation
\begin{eqnarray}  \displaystyle{
[\mu x^i_{C.M.}  \:  ,   \:   p_i] = i \mu
}  \end{eqnarray}
indicates that the mass $\mu$ must be included to close the algebra. The mass
commutes with the other generators, so it is a central charge. We leave it as an
exercise for the reader to compute the remaining commutation relations.
\par Now consider the Poincar\'e generators of $D$ dimensional space-time:
\begin{eqnarray}  \nonumber \displaystyle{
\begin{array}{cl}
P_i \:\:\:\:\:\:\:\:\:\:  & \:\:\:\:\:\:\:\:\:\:
\mathrm{transverse\:translations} \\
P_+=H  \:\:\:\:\:\:\:\:\:\: &  \:\:\:\:\:\:\:\:\:\:
X^+ \: \mathrm{translations} \\
P_-   \:\:\:\:\:\:\:\:\:\:   &    \:\:\:\:\:\:\:\:\:\:
X^- \: \mathrm{translations} \\
L_{ij} \:\:\:\:\:\:\:\:\:\:  &  \:\:\:\:\:\:\:\:\:\:
\mathrm{transverse\:rotations}  \\
L_{i z}    \:\:\:\:\:\:\:\:\:\:  &   \:\:\:\:\:\:\:\:\:\:
{\mathrm{rotations\:in\:}}  (X^i, z) \: \mathrm{plane}  \\
K_{0 z}  \:\:\:\:\:\:\:\:\:\:  &  \:\:\:\:\:\:\:\:\:\:
{\mathrm{Lorentz\:boosts\:along\:}}  z  \\
K_{0 i}  \:\:\:\:\:\:\:\:\:\:  &   \:\:\:\:\:\:\:\:\:\:
{\mathrm{Lorentz\:boosts\:along\:}}  X^i \\
\end{array}
}  \end{eqnarray}
Now define the following correspondence:
\begin{eqnarray}  \nonumber \displaystyle{
\begin{array}{ccc}
P_i  \:\: & \:\: \longleftrightarrow \:\:  &   \:\: p_i \\
P_+=H  \:\: & \:\: \longleftrightarrow \:\:  &   \:\: h \\
P_-  \:\: & \:\: \longleftrightarrow \:\:  &   \:\: \mu \\
L_{ij}  \:\: & \:\: \longleftrightarrow \:\:  &   \:\: l_{ij} \\
B_i = \frac{K_{0i}+ L_{0i}}{\sqrt{2}}   \:\:  & \:\: \longleftrightarrow \:\:\
&   \:\:   \mu x^i_{C.M.}
\end{array}
}  \end{eqnarray}
Using the standard Poincar\'e commutation relations one finds an exact
isomorphism between the Galilean group and the Poincar\'e subgroup generated
by $P_i$, $H$, $P_-$, $L_i$ and $B_i$. Thus it follows that relativistic
physics
in the light cone frame is Galilean invariant and must have all the properties
which follow from this symmetry. For example the Hamiltonian must have the form
\begin{eqnarray}  \displaystyle{
H= \frac{P_i^2}{2P_-} + E_{\mathrm{internal}}
} \end{eqnarray}
where $E_{\mathrm{internal}}$ is Galilean invariant.
\par The Poincar\'e generator $K_{0 z}$ is not part of the Galilean subgroup,
but it gives important information about $E_{\mathrm{internal}}$.
Using the commutation relations
\begin{eqnarray}  \displaystyle{
[K_{0z}, H] = H
}  \end{eqnarray}
\begin{eqnarray}  \displaystyle{
[K_{0z}, P_-] = - P_-
}  \end{eqnarray}
we see that the product $HP_-$ is boost invariant. It follows that
$E_{\mathrm{internal}}$ has the form $M^2/(2P_-)$ with $M^2$ being invariant
under Galilean boosts and Lorentz transformations. In other words $H$ must
scale like $1/P_-$ under a rescaling of all $P_-$ in the system.
It is interesting to think of rescaling the $P_-$ axis as a kind of scale
transformation. The invariance of physics under longitudinal boosts is
understood as the existence of a renormalization group fixed point
 \cite{7} .
\par Let us now consider the formulation of quantum field theory in the light
cone frame. Let $\phi$ be a scalar field with action
\begin{eqnarray}  \displaystyle{
A= \int d^D X \: \left\{
\frac{\partial_\mu \phi \partial^\mu \phi}{2}  -  \frac{m^2 \phi^2}{2}  -
\lambda \phi^3   \right\}
}  \end{eqnarray}
In terms of light cone coordinates this becomes
\begin{eqnarray}  \displaystyle{
A= \int dX^+ \: {\cal L }
}  \end{eqnarray}
\begin{eqnarray} \label{14} \displaystyle{
{\cal L} = \int dX^- \: dX^i \: \left\{ \dot{\phi}  \: \partial_- \phi -
\frac{(\partial_i \phi)^2}{2}  - m^2 \phi^2 - \lambda \phi^3        \right\}
}  \end{eqnarray}
where the notation $\displaystyle{\dot{\phi} = \partial_+ \phi }$ has been
used. We can now identify the canonical momentum $\pi$ conjugate to $\phi$
\begin{eqnarray}  \displaystyle{
\pi = \frac{\delta {\cal L}}{\delta \dot{\phi}}  = \partial_- \phi
}  \end{eqnarray}
from which we deduce the equal time commutation relations
\begin{eqnarray}  \displaystyle{
[\phi(X^i, X^-), \phi'(Y^i, Y^-)] = i \: \delta (X^i-Y^i) \: \delta(X^- -Y^-)
}  \end{eqnarray}

We may Fourier transform the field $\phi$ with respect to $X^-$:
\begin{eqnarray}  \nonumber \displaystyle{
\phi(X^i, X^-) = \int_0^\infty dk_- \: \frac{\phi(X^i,k_-)}{\sqrt{2 \pi 
|k_-|}}
e^{ik_- X^-}  +
} \end{eqnarray}
\begin{eqnarray} \label{17a} \displaystyle{
+ \frac{\phi^*(X^i,k_-)}{\sqrt{ 2 \pi |k_-|}} e^{-ik_- X^-}
}  \end{eqnarray}
The Fourier coefficients $\phi$, $\phi^*$ have the non relativistic
commutation relations
\begin{eqnarray}  \label{18a} \displaystyle{
[\phi(X^i, k_-), \phi(Y^i, l_-)] = 0
}  \end{eqnarray}
\begin{eqnarray}  \label{19a} \displaystyle{
[\phi^*(X^i, k_-), \phi^*(Y^i, l_-)] =  0
}  \end{eqnarray}
\begin{eqnarray}  \label{20a} \displaystyle{
[\phi(X^i, k_-), \phi^*(Y^i, l_-)] =  \delta (X^i-Y^i) \: \delta(k_- -l_-)
}  \end{eqnarray}
The Hamiltonian following from (\ref{14}) has the form
\begin{eqnarray} \label{19} \displaystyle{
H=\int dk_- dX^i \frac{\nabla \phi^* \cdot \nabla \phi + m^2 \phi^*
\phi}{2 k_-}  +  H_{\mathrm{interaction}}
}  \end{eqnarray}
The first term in $H$ may be compared with the hamiltonian of a
system of free particles in nonrelativistic physics. Let $\psi_k (X)$
be the second quantized Schr\"odinger field for the $k$th type
of particle. Then
\begin{eqnarray}  \displaystyle{
H_{\mathrm{N.\:R.}} = \sum_k \frac{\nabla \psi_k^\dagger (x) \:
\nabla \psi_k (x) }{2 \mu_k} + \frac{U_k}{2 \mu_k} \psi^\dagger \psi
}  \end{eqnarray}
Evidently, the Hamiltonian in eq.~(\ref{19}) has the nonrelativistic
form, except that the discrete sum over particle type is replaced
by an integral over $k_-$.
\par A very important point to notice is that the $k_-$ integration
runs only over non negative $k_-$. There are no quanta with $k_-<0$.
This of course is analogous to the positivity of mass in 
non relativistic quantum mechanics. 
\par The interaction term $H_{\mathrm{interaction}}$ has the form
\begin{eqnarray}  \displaystyle{
H_{\mathrm{interaction}} \propto \int dk_-  \: dl_- \: 
\frac{\phi^\dagger (k_-)}{\sqrt{|k_-|}}
\frac{\phi^\dagger (l_-)}{\sqrt{|l_-|}} 
\frac{\phi(k_- +l_-)}{\sqrt{|k_-+ l_-|}} + {\mathrm{c.\:c.}}
}  \end{eqnarray}
The important thing to note is that the value of $k_-$ is conserved
by the interaction. This, together with the positivity of $k_-$,
insures that there are no terms like $\phi^\dagger \phi^\dagger
\phi^\dagger$ which create quanta from the Fock space vacuum.
For this reason the Fock space vacuum is the true vacuum in the
Fock space quantization.
\par Perturbative processes induced by $H_{\mathrm{interaction}}$
are generated by vertices which allow one particle to split into two,
or the reverse, conserving $k_-$.
\subsection{DLCQ}
\par Matrix theory is based on a form of light cone quantization
called ``discrete light cone quantization''
 \cite{8}. To define DLCQ, the
light like coordinate $X^-$ is compactified to a circle of
circumference $2\pi R$. The effect of this compactification is to
discretize the spectrum of $P_-$:
\begin{eqnarray}  \displaystyle{
P_-= \frac{N}{R}
}  \end{eqnarray}
where $N$ is a non negative integer. Since $P_-$ is conserved,
the system splits up into an infinite number of superselection
sectors characterized by $N$.
\par Equations (\ref{17a}), (\ref{18a}), (\ref{19a}), (\ref{20a})
are replaced by
\begin{eqnarray}  \displaystyle{
\phi = \phi_0 + \sum_{n=1}^\infty \frac{\phi_n(X^i)}{\sqrt{2 \pi n}}
e^{in \frac{X^-}{R}} + {\mathrm{c.\:c.}}
} \end{eqnarray}
\begin{eqnarray} \displaystyle{
[\phi^\dagger_n ,  \phi^\dagger_m] = [\phi_n, \phi_m]=0
}  \end{eqnarray}
\begin{eqnarray}  \displaystyle{
[\phi^\dagger_n(X^i), \phi_m(Y^i)]= \delta_{nm} \delta(X^i-Y^i)
}  \end{eqnarray}
where $\phi_0$ is the mode of $\phi$ with $k_-=0$.
The Hamiltonian becomes a discrete series of terms
\begin{eqnarray} \nonumber  \displaystyle{
H=R \left\{ \sum_{n=1}^\infty \frac{\nabla \phi_n^* \cdot
\nabla \phi_n + m^2}{2n} \: +
\right.
}  \end{eqnarray}
\begin{eqnarray}  \displaystyle{
\left. + {\mathrm \:  const. \:}\sum_{n,m=0}^\infty 
\frac{\phi^\dagger (n, \: X^i)}{\sqrt{n}}
\frac{\phi^\dagger (m, \: X^i)}{\sqrt{m}} 
\frac{\phi(n+m, \: X^i)}{\sqrt{n+m}}  
\right\} + {\mathrm c.\:c. }
}  \end{eqnarray}
The ``zero mode'' $\phi_0$ is non dynamical and can
be integrated out, giving rise to new terms in $H$.
These new terms conserve $P_-$ and preserve the
Galilean symmetry (provided that $N$ is conserved).  Other than that, they
may be of
arbitrary complexity. For example $\phi^\dagger
\phi \phi \phi$, $\phi^\dagger \phi^\dagger \phi
\phi$, $\ldots$ terms may be induced.
\par Quantum mechanics within a given $N$ sector
is much simpler than in the uncompactified theory.
For example, in the sector $N=1$ nothing interesting
can happen. The spectrum is a single particle which
can not split into constituents. For $N=2$ the Hilbert
space is a sum of one particle states with two units
of $P_-$ and two particle states, each particle
carrying $N=1$. The only processes which can occur
are splitting of the 2-unit particle into two
one-unit particles and scattering of the 2-unit particles.
However, as $N$ grows the number of allowed
processes grows.
\par Physical applications require that the limit $N
\rightarrow \infty$ be taken. To see this, we need
only note that the quantum number $N$ is given by
 \begin{eqnarray}  \displaystyle{
N=P_- R
}  \end{eqnarray}
so that if we fix the physical component of momentum
and let the radius $R$ tend to infinity, $N$ also becomes
infinite.
\par Any attempt to use DLCQ as a numerical approximation
scheme should begin with an estimate of how large $N$
needs to be in order to achieve a given degree of accuracy
for a given problem
 \cite{9}. A rough estimate can be obtained
from geometrical considerations. Consider a system with a
given mass $M$ whose largest spatial dimension is of order
$\rho$. Assume the system is at rest in the transverse plane.
That is,
\begin{eqnarray}  \displaystyle{
P_i=0
}  \end{eqnarray}
\par Let us also boost the system along the $z$ axis until it is at rest
with $P_z=0$. Its momentum vector is purely timelike with $P_-=
P_+=M$. In this frame the longitudinal size $\Delta z$ of the system is no
larger than $\rho$. Obviously for DLCQ to give a good approximation the
size of the system should not exceed the
compactification scale $R$. Thus, the condition for a good
approximation is
\begin{eqnarray}  \displaystyle{
\rho   {\
\lower-1.2pt\vbox{\hbox{\rlap{$<$}\lower5pt\vbox{\hbox{$\sim$}}}}\ } R
}  \end{eqnarray}
Multiplying by $P_-=M$ gives
\begin{eqnarray}  \displaystyle{
M{\ \lower-1.2pt\vbox{\hbox{\rlap{$<$}\lower5pt\vbox{\hbox{$\sim$}}}}\ } R
P_-
}  \end{eqnarray}
But $RP_-$ is $N$, so that the condition is
\begin{eqnarray}  \displaystyle{
N {\ \lower-1.2pt\vbox{\hbox{\rlap{$>$}\lower5pt\vbox{\hbox{$\sim$}}}}\ } M
\rho
}  \end{eqnarray}
Later we will use this condition in studying black holes.

\subsection{Another view of DLCQ}
The light cone frame can be characterized by its metric
\begin{eqnarray}  \displaystyle{
ds^2= dX^+ dX^- - (dX^i)^2
}  \end{eqnarray}
In order to resolve some of the ambiguities inherent in lightlike
compactification, it is useful to introduce a frame in which the
metric has the ``regularized'' form
 \cite{10} 
\begin{eqnarray}  \displaystyle{
ds^2= d \tilde{X}^+ d \tilde{X}^- - \epsilon^2 (d \tilde{X}^-)^2
- (dX^i)^2
}  \end{eqnarray}
The limit $\epsilon \rightarrow 0$ defines the light cone frame,
but now the direction $\tilde{X}^-$ is a true spacelike direction.
Compactification of $\tilde{X}^-$ on a spacelike circle involves
only standard procedures. Note, however, that if $X^-$ is
periodic with radius $R$ the proper size of the spacelike circle is
\begin{eqnarray}  \displaystyle{
R_c=\epsilon R
}  \end{eqnarray}

Evidently DLCQ may be interpreted as the limit of spacelike
compactification in which the compactification size shrinks to
zero. However that is not all there is to it. Consider the sector of
DLCQ with $P_- = N/R$. This may now be interpreted in the
spacelike compactified theory as the sector with spacelike
momentum
\begin{eqnarray}  \displaystyle{
\tilde{P}_- = \frac{N}{R_c} = \frac{1}{\epsilon} \frac{N}{R}
}  \end{eqnarray}
In other words DLCQ in sector $N$ is obtained by the following
procedure:
\begin{enumerate}
\item{compactify on a spacelike circle of size $R_c$;}
\item{consider the sector with $N$ units of quantized
spacelike momentum;}
\item{holding $N$ fixed let $R_c \rightarrow 0$; in this case 
the spacelike $\tilde{P}$ tends to $\infty$. }
\item{Now boost the system back to the original frame in which the
compactification radius is $R$. The boost factor or time dilation factor is
$R_c/R$ and
therefore goes to infinity}. The time dilation factor also implies a
rescaling of the
hamiltonian
by a factor  $R/R_c$.
\end{enumerate}

Thinking about DLCQ from this point of view illuminates the problems
associated with the
zero modes $\phi_0$. The zero mode sector of a compactified theory defines
a field theory in a
lower dimension but with a coupling which blows up as $R_c \to \infty$.
Thus integrating the zero
modes out of a field theory is nontrivial and involves the solution to a
strong coupling field
theory which can not be obtained by perturbative means
 \cite{11}.

One may wonder how it is possible to retrieve all of the physics of the
uncompactified
theory from the theory on the vanishingly small circle of size $R_c$. The
answer is Lorentz
contraction. Suppose we focus on an object of size $b$ and mass $M$. Let us
boost this object so
that  its momentum in the $z$ direction is $N/R_c$. Its longitudinal size
is reduced to $b'= b 
{MR_c  \over N}$. Evidentally as $N$ gets large the system under
consideration becomes much
smaller than $R_c$. In the frame of the moving system the compactification
radius becomes so big
that it has no effect of the system. Thus we see that the system
compactified  on $R_c$ contains
all the information about the uncompactified theory.

\subsection{Supergalilean symmetry}
\par For supersymmetric theories, the Galilean symmetry can be augmented
with anticommuting
spinorial supercharges to form the supergalilean group. Let us focus on the
11-dimensional case
which  will be of particular interest in the following chapters. We begin
with the rotation group $O(9)$ representing rotations in the
9-dimensional transverse space. The spinor representation of $O(9)$
is real and 16 dimensional. Spinors are labeled $\theta_\alpha$ with
$\alpha=(1, \ldots, 16)$. The Dirac matrices $\gamma_{\alpha, \beta}^i$
are real.
\par The superalgebra of 11 dimensional supergravity has 32 real
supercharges which can be
separated into two sets of 16, namely
$Q_\alpha$ and $q_\alpha$. The $Q_\alpha$ anticommute to give
the light cone Hamiltonian
\begin{eqnarray}  \displaystyle{
[Q_\alpha, Q_\beta]_+ = \delta_{\alpha \beta} H = \delta_{\alpha \beta}
P_+
}  \end{eqnarray}
Similarly, the $q_\alpha$ close on $P_-$:
\begin{eqnarray}  \displaystyle{
[q_\alpha, q_\beta]_+ = \delta_{\alpha \beta} P_-
}  \end{eqnarray}
and the mixed anticommutation relations are
\begin{eqnarray}  \displaystyle{
[Q_\alpha, q_\beta]_+ = \gamma_{\alpha \beta}^i P_i
}  \end{eqnarray}
\par A simple one-particle representation of the supergalilean group
may be constructed as follows. The particle states are labeled by
$P_-$, $P_i$ and 16 anticommuting coordinates $\theta_\alpha$. The
supercharges are
\begin{eqnarray}  \displaystyle{
Q_{\alpha}= \frac{ P_i \gamma_{\alpha \beta}^i \theta_{\beta}}{\sqrt{2
P_-}}
}  \end{eqnarray}
\begin{eqnarray}  \displaystyle{
q_{\alpha}=\theta_{\alpha} \sqrt{P_-}
}  \end{eqnarray}
The Hamiltonian is easily seen to be
\begin{eqnarray}  \displaystyle{
H= \frac{P_i^2}{2P_-}
}  \end{eqnarray}

\section{Lecture 2 (Matrix theory)}
\subsection{Type IIA string theory and M theory}

One of the big string theory surprises of the last few years
was the discovery that string theory implies the existence
of an 11-dimensional theory which itself is not a string theory.
This theory, called M-theory
\cite{5}, is the limit of 10-dimensional
IIA string theory in which the string coupling $g_{st}$ tends to
infinity .

There were a number of hints that an 11-D theory might underlie
IIA string theory. The first was the existence of the dilaton field
in the low energy action. Typically when a gravitational theory
is compactified (let us call the compact dimension $z$), the
component of the metric $g_{zz}$ behaves like a scalar field
in the lower dimensional theory. Furthermore, it enters the
action of a lower dimensional theory in the same way that the
dilaton does. This suggests that the dilaton is really the local
compactification radius of the $z$ direction. The second piece of
evidence was the existence of an abelian gauge field in the
10-dimensional IIA theory. Generally speaking, all other gauge
fields that occur in string theory can be thought of as
Kaluza-Klein gauge fields, either before or after some kind of
duality transformation. The so-called RR gauge boson of type
IIA theory was not obviously a K-K object. However, with the
interpretation of IIA as the K-K compactification of M-theory,
the R-R gauge field found its natural interpretation in terms of the
$g_{\mu z}$
components of the 11-D metric.

What was missing from perturbative string theory were the
charged sources of the R-R field. In Kaluza Klein theory
these sources are the quantized $z$-components of momenta
of particles moving in 11 dimensions. However the KK charges
were soon located in the form of Polchinski's D-branes
 \cite{12}.
In particular, the only objects carrying the appropriate charge are
the D0-branes, particle-like objects on which strings may end.
One of the main things we will need to know about these objects is their
mass. Let us begin with a massless 11-dimensional particle with
an unit of Kaluza-Klein momentum. The mass of this particle in
the lower dimensional theory is given by
\begin{eqnarray} \displaystyle{
M_0=\frac{1}{R_C}
}\end{eqnarray}
where $R_C$ is the compactification radius of the $z$ direction;
in other words it is the background value of the dilaton field.

On the other hand, D0-branes have a mass which can be computed in
string theory and is given
by
 \cite{12} 
\begin{eqnarray} \displaystyle{
M_0= \frac{1}{g_{st} l_{st}}
}\end{eqnarray}
where $l_{st}$ is the basic length scale in string theory.

Evidently, the KK interpretation requires
\begin{eqnarray} \label{qqq} \displaystyle{
R_C= g_{st} l_{st}
}\end{eqnarray}
This explains why the KK charges do not appear in perturbative
string theory. Their masses diverge as $g_{st} \rightarrow
0$ and they decouple from perturbation theory. In this limit
the compactification scale $R_C \rightarrow 0$ and becomes
invisible.

On the other hand, the limit $g_{st} \rightarrow \infty$ is
characterized by very light D0-branes and by the
compactification radius becoming infinite. This limit is called
M-theory.

What do we know about M-theory? First of all it is a 11-D
gravitational theory with a Planck length $l_{11}$. The
gravitational coupling in 11-D is
\begin{eqnarray} \displaystyle{
G_{11} = l_{11}^{\: 9}
}\end{eqnarray}

Furthermore, since IIA string theory is supersymmetric with 32
real supercharges, so must be M-theory. This fits conveniently
with the known properties of 11-D supergravity. Thus we
assume that low energy M-theory is governed by 11-D
supergravity.

Besides the gravitational field $g_{\mu \nu}$, 11-D
supergravity has another bosonic field, a 3-form $A_{\mu \nu
\rho}$ and a fermionic gravitino $\psi$.
In addition to the quanta of these fields, M-theory is postulated
to have two other forms of matter. Both are associated with the
form field $A_{\mu \nu \rho}$. Just as a one-form (a gauge field)
couples to the world lines of particles, a 3-form can couple to the
3-dimensional world volume swept out by a membrane or
2-brane. Since in M-theory the only natural length scale is the 11-D Planck
length,
the energy of the membrane is governed
by a tension (energy/area) of order
\begin{eqnarray}  \displaystyle{
T_2 = \frac{1}{l_{11}^{\:3}}
}  \end{eqnarray}

The other form of matter is related to $A_{\mu \nu \rho}$ by
an analogue of electric/magnetic duality. The second form of
matter is 5-branes which sweep out 6-dimensional world
volumes.

It is now possible to understand the origin of strings in
M-theory when one of the space coordinates $z$ is
compactified. Consider a membrane with the topology of
a torus. Its spatial volume is parametrized by two angles,
$\theta_1$ and $\theta_2$. Let us suppose it is embedded in
11-D so that one cycle, say $\theta_1$, is wrapped
around the compact coordinate $z$:
\begin{eqnarray}  \displaystyle{
\theta_1 = \frac{z}{R_C}
}  \end{eqnarray}
The other spatial directions $X_1, \ldots, X_9$ are functions of
$\theta_2$.

Such a configuration defines a string in the 9 noncompact
space dimensions
\begin{eqnarray}  \displaystyle{
X^i = X^i(\theta_2)  \: \: \: \: \: \: \: \: \: \: \: \:  i=1, \ldots, 9
}  \end{eqnarray}
The tension of the string is easily computed since the energy
per unit area of the membrane is
\begin{eqnarray}  \displaystyle{
T_2= \frac{1}{l_{11}^{\: 3}}
}  \end{eqnarray}
The energy per unit length of the string is
\begin{eqnarray}  \displaystyle{
T_1= \frac{R}{l_{11}^{\: 3}}
}  \end{eqnarray}
On the other hand, the string tension and string length scale
are related by
\begin{eqnarray}  \displaystyle{
T_1= \frac{1}{l_s^{\: 2}}
}  \end{eqnarray}
Thus
\begin{eqnarray} \label{*} \displaystyle{
l_s^{\: 2} = \frac{l_{11}^{\: 3}}{R_C}
}  \end{eqnarray}
Combining this with equation (\ref{qqq}) we find
\begin{eqnarray}  \label{**} \displaystyle{
g_{st}^2 = \frac{R_C^2}{l_{st}^2} = \frac{R_C^3}{l_{11}^{\: 3}}
}  \end{eqnarray}
Equations (\ref{*}) and (\ref{**}) define the connection between
string theoretic and M-theoretic quantities.

\subsection{DLCQ of M-theory}
\par Let us now construct the DLCQ of M-theory. According to
sect.~1.2 we begin by
compactifying M-theory on a space-like circle of radius $R_c$.
But we have just seen that this is equivalent to type IIA string
theory on a circle of size
\begin{eqnarray}  \displaystyle{
R_c^{\: 3}= g_{st}^2 l_{11}^3
}  \end{eqnarray}

Next we focus on the sector with spacelike momentum $N/R_c$.
In IIA language this means the sector with $N$ units of
D0-brane charge. Finally, holding $N$ fixed we pass to the
limit $R_c \rightarrow 0$ or equivalently $g_{st} \rightarrow 0$.
In other words we pass to the limit of infinitely weakly coupled
string theory. One subtle point is that in passing to the limit we
do not hold the string scale fixed. We are interested in phenomena on the scale
of $l_{11}$ so it is appropriate to keep
this scale fixed. The string scale goes to infinity according to
eq.~(\ref{*}).

Note that even in 11 dimensional Planck units the (10
dimensional) masses of the D0-branes tend to infinity as
$R_c \rightarrow 0$. This means that the D0-branes become
nonrelativistic in the 10-D sense. It is therefore natural that
they are described by Galilean quantum mechanics in 9
spatial dimensions. This Galilean symmetry is of course exactly
the Galilean symmetry of DLCQ.

Thus we come to the remarkable conclusion that M-theory is
equivalent to the $N \rightarrow \infty$ limit of the nonrelativistic
quantum mechanics of $N$ D0-branes in weakly coupled limit of IIA string
theory.

\subsection{D0-brane quantum mechanics}
Fortunately the quantum mechanics of $N$ nonrelativistic
D0-branes has been worked out in \cite{13}, \cite{14}. 

Let us begin with the theory of a single D0-brane. The particle
is characterized by a transverse location $X^i$ and a 10-D
mass $\frac{1}{R_c}$. The D0-brane Lagrangian is
\begin{eqnarray}  \displaystyle{
{\cal L} = \frac{\epsilon}{2 R_c} (\dot{X}^i)^2
}  \end{eqnarray}
where the factor ${\epsilon}$ represents the time dilation 
factor from 10-D time to 11-D $X^+$ described in sect.~2. 

The Hamiltonian of the system satisfies
\begin{eqnarray}  \displaystyle{
H= \frac{1}{2 \epsilon} R_c P_i^2 = \frac{1}{2}  R P_i^2
}  \end{eqnarray}
In order to make the system supersymmetric we add 16 fermionic zero mode
coordinates $\theta$ as in sect.~2. The
D0-brane then has the spectrum and quantum numbers of 11-D supergraviton.

Next, let us consider $N$ D0-branes. We begin with $N$
positions and fermionic partners $X^i_a$, $\theta_a$
($a=1, \ldots N$). The obvious action is
\begin{eqnarray}  \displaystyle{
{\cal L} = \sum_a{ \left[ \frac{1}{2 R_c} \dot{X}_a^{\: 2} +
\dot{\theta}_{\dot{a}}  \theta_a  \right]}
}  \end{eqnarray}

To describe the interactions between D0-branes, we recall that in
IIA string theory a D0-brane is the Dirichlet boundary where a
string may end. Thus we may consider strings which connect
the D0-branes ($a$, $b$). Since the string tension is given by
$\frac{R_c}{l_{11}^3}$, the minimum energy of such a string is
\begin{eqnarray}  \displaystyle{
E(a,b) = \frac{|X_a -X_b| \: R_c}{l_{11}^3}
}  \end{eqnarray}

Consider an excited string state with additional energy
\begin{eqnarray}  \displaystyle{
\Delta E \sim \frac{1}{l_s} = \left( \frac{R_c}{l_{11}^3}
\right)^{1/2}
}  \end{eqnarray}
Note that in the limit $R_c \rightarrow 0$ the excitation energy
of a string becomes infinitely bigger than the minimum 
excitation energy
$E(a,b)$. Thus one may expect that in the limit $R_c \rightarrow
0$ only the lightest strings connecting ($a$, $b$) are relevant.
\par In string theory, strings are similar to quanta of a field that
can be created and annihilated. It is therefore plausible to
describe these lightest states by field operators. A 
systematic analysis
shows that the strings can be either bosons or fermions and
that the bosonic strings carry a polarization index $i$ which
transforms as a vector under 9-dim.~rotations. Similarly, the
fermions form spinors. Accordingly we label the string fields
$X^i_{ab} , \:  \theta_{ab}$.

If we relabel the coordinate locations and the thetas by
\begin{eqnarray}  \displaystyle{
X_a \rightarrow X_{aa}
}  \end{eqnarray}
\begin{eqnarray}  \displaystyle{
\theta_a \rightarrow \theta_{aa}
}  \end{eqnarray}
we see that the degrees of freedom can be assembled into
$N \times N$ matrix degrees of freedom that include positions
(diagonal elements) and stretched strings as well as their
fermionic superpartners.

The action for the matrix degrees of freedom can be derived
from IIA string theory \cite{14}. In this lecture we will not carry 
out the derivation but merely write down the result:
\begin{eqnarray}  \label{action} \displaystyle{
{\cal L} = Tr \left[ \frac{1}{2 R_c} \dot{X}^2 +
\frac{R_c}{4 \: l_{11}^6}  [X^i, X^j]^2
+ \mathrm{fermions}  \right]
}  \end{eqnarray}
The action (\ref{action}) has a symmetry that might appear
accidental from the incomplete discussion in this paper. The
symmetry is $U(N)$ where the action of the symmetry group is
\begin{eqnarray}  \displaystyle{
X \rightarrow U^{\dagger} X U
}  \end{eqnarray}
\begin{eqnarray}  \displaystyle{
\theta \rightarrow U^{\dagger} \theta U
}  \end{eqnarray}
This symmetry has as its origin the underlying gauge symmetry
of string theory. It is the same symmetry that leads to gauge
simmetries for open string theories and it should be treated as a
gauge symmetry. To make the gauge invariance manifest we may 
introduce a vector potential $A_0$ in the adjoint of $U(N)$.
The time derivatives in the action can be replaced by
covariant derivatives:
\begin{eqnarray}  \displaystyle{
{d{x} \over dt} \rightarrow {d{x} \over dt} + [A_0, X]
}  \end{eqnarray}
The only effect of including $A_0$ into the action is to introduce
a Gauss law constraint which requires the generators of $U(N)$
to vanish on physical states. In other words only the states which
are invariant under the $U(N)$ symmetry are allowed in the
physical spectrum.

The Hamiltonian resulting from eq.~(\ref{action}) is obtained
straightforwardly. The momentum conjugate to $X^i_{ab}$ is
\begin{eqnarray}  \displaystyle{
P_i^{ab} = \frac{1}{R_c} \dot{X}^i_{ab}
}  \end{eqnarray}
The Hamiltonian (after rescaling by the time dilation factor $R/R_c$ ) is
\begin{eqnarray}  \label{70} \displaystyle{
H= R \: Tr \left[ \frac{1}{2} P_i^2 - \frac{1}{4 \: l_{11}^6} 
[X^i, X^j]^2 
+ {\mathrm fermions}  \right]
}  \end{eqnarray}
Thus we arrive at the precise form of the matrix theory
conjecture:
\begin{enumerate}
\item{Whatever M-theory is, it may be formulated in a world with
a compact lightlike direction and quantized according to the
rules of DLCQ;}
\item{In the sector of the theory with $P_-= \frac{N}{R}$ the
theory is exactly described by the hamiltonian (\ref{70})
describing supersymmetric quantum mechanics with 16 real
supersymmetries and $U(N)$ symmetry.}
\end{enumerate}
\par The supergalilean symmetry of eq.~(\ref{70}) required by
the DLCQ interpretation is easy to prove. For example a
Galilean boost acts on $\Pi$ according to
\begin{eqnarray}  \displaystyle{
\Pi \rightarrow \Pi + \frac{1}{R} {v} \otimes I
}  \end{eqnarray}
where ${v}$ is a parameter and $I$ is the $N \times N$
identity matrix. Note also that the quartic potential is invariant
under translations
\begin{eqnarray}  \displaystyle{
X^i \rightarrow X^i + a^i \otimes I
}  \end{eqnarray}

\subsection{The supergraviton}
M-theory must contain in its spectrum the massless graviton and
its various superpartners. Let us consider how the spectrum of
single and multiple supergravitons arise. We begin with $N=1$ in
which case the Hamiltonian is
\begin{eqnarray}  \displaystyle{
H= \frac{R}{2} P_i^2
}  \end{eqnarray}
This is the correct expression in DLCQ for any massless particle
with $P_-= 1/R$. In this case the $\theta$ do not appear in $H$
but they do provide the spin degrees of freedom. The 16
$\theta$ can be grouped into 8 complex fermionic creation
operators and their conjugates. Therefore the algebra of the
$\theta$ can be realized on a space of dimensionality $2^8$.
The 256 states consist of 128 bosons and 128 fermions. It is
found that the bosons include 44 gravitons which transform
under $O(9)$ as a symmetric traceless tensor and 84 objects
transforming as an antisymmetric tensor with three indices.
This of course is exactly the bosonic content of the 11-D
supergraviton. The 128 fermions are the gravitinos.

Now let us consider the general case of arbitrary $N$. To do
this we divide the degrees of freedom into $U(1)$ and
$SU(N)$ degrees of freedom by separating off the trace.
This is identical to a separation into center of mass and relative
motions.
\begin{eqnarray}  \displaystyle{
X_{c.m.} = \frac{1}{N} \: Tr X
}  \end{eqnarray}
\begin{eqnarray}  \displaystyle{
\theta_{c.m.} = \frac{1}{\sqrt{N}} \: Tr \theta
}  \end{eqnarray}
\begin{eqnarray}  \displaystyle{
P_{c.m.} = Tr P
}  \end{eqnarray}
\begin{eqnarray}  \displaystyle{
X-X_{c.m.} = X_{rel}
}  \end{eqnarray}
\begin{eqnarray}  \displaystyle{
\theta - \frac{1}{\sqrt{N}} \theta_{c.m.} = \theta_{rel}
}  \end{eqnarray}
\begin{eqnarray}  \displaystyle{
P- \frac{P_{c.m.}}{N} = P_{rel}
}  \end{eqnarray}

From the form of $H$ we see that the c.~m.~degrees of freedom
are completely decoupled from the relative variables. The
hamiltonian may be written as
\begin{eqnarray}  \displaystyle{
H= \frac{R P_{c.m.}^2}{2N} + H_{rel}
}  \end{eqnarray}
where $H_{rel}$
 is the Hamiltonian with $SU(N)$ symmetry for the relative
variables.

Let us suppose that $H_{rel}$ has a normalizable
supersymmetric ground state. Supersymmetry requires the
eigenvalue of $H_{rel}$ to vanish. Thus for the ground state of
the relative motion the Hamiltonian reduces to
\begin{eqnarray}  \displaystyle{
H= \frac{R P_{c.m.}^2}{2N} = \frac{P_{c.m.}^2}{2 P_-}
}  \end{eqnarray}
which again has the right form for a massless particle. Since
$\theta_{c.m.}$ does not enter in $H$ it again provides the spin
degeneracy of the supergraviton.

The existence of a normalizable ground state for the
c.~m.~variables is not at all trivial. To see why, let us consider the
potential energy $- Tr \: R [X^i, X^j]^2$. This term provides a
potential which confines the $X_{c.m.}$ to the region near the
origin. However there are ``flat directions'' along which they can
escape. For example consider matrices $X^i$ which all commute.
To understand the meaning of such configurations we note
that the mutual commutation implies that the $X$ can be
simultaneously diagonalized. If we do so, then the configuration
can be labeled by $N$ 9-vectors $X^i_a$ representing the
location of the $N$ D0-branes. In other words the flat directions could allow
the D0-branes to leak away from the bound state.
As we shall argue, the flat directions are not lifted by quantum
corrections, so the likelyhood remains that the bound configurations
representing a
supergraviton of $N$ units
of $P_-$ may be unstable.

Thus, it is very significant that the existence of a normalizable
ground state for the relative variables has recently been
proved by Sethi and Stern and by Yi. The theorem is 
surprisingly difficult to prove and in fact depends crucially 
on the very special properties of the system, including its 
16 supersimmetries. 

Having proved the existence of single supergraviton states, we
now turn to multigraviton configurations. In particular, the
theory must admit free multigraviton configurations in the limit
of large separation. Thus let us try to construct a two supergraviton
state. To do this we
temporarily restrict attention
to $N \times N$ matrices $X$, $\theta$ which have the form of
$N_1 \times N_1 + N_2 \times N_2$ block diagonal matrices.
Let us call the elements in the upper block $X^i_{ab}$, $\theta_{ab}$ and
the ones
in the lower block $Y^i_{ab}$, $\phi_{\alpha \beta}$. The off diagonal
blocks are
temporarily frozen to zero. If we now substitute such matrices into the
action (eq.~(\ref{action}))
we find that the system separates into two uncoupled matrix models,
one with $N_1 \times N_1$ matrices and the other with $N_2 \times N_2$
matrices. Quantizing
this system naturally provides a pair of supergravitons with momenta
$N_1/R$ and $N_2/R$.

Freezing the off diagonal elements is, of course, not a legitimate thing to
do. But let us
assume
that the center of masses of the two gravitons are very distant:
\begin{eqnarray}  \displaystyle{
| \frac{1}{N_1} Tr X - \frac{1}{N_2} Tr Y | = |X_{c.m.} - Y_{c.m.} |
 \gg  l_{11}
}  \end{eqnarray}
Now consider the equation of motion for an off diagonal element which we
call $w$. The
action for $w$ in the background of $X$ and $Y$ has the form
\begin{eqnarray}  \displaystyle{
{\cal L}_z = \frac{1}{2 R} \dot{w}^2 - \frac{R |X_Y|^2}{l_{11}^6} w^2
}  \end{eqnarray}
where the second term arises from the commutator term in (\ref{action}).
The off diagonal elements behave like oscillators with frequency
\begin{eqnarray}  \label{fff3} \displaystyle{
\omega = \frac{R |X-Y|}{l_{11}^3}
}  \end{eqnarray}
Note that this is exactly the energy of a string stretched from $X$ to $Y$.
This was to be
expected since the off diagonal elements of the matrices correspond to string field
operators.

The important thing to note about eq.~(\ref{fff3}) is that as $|X-Y|$ grows
the frequency of the
mode $w$ tends to infinity. One can therefore expect that the off diagonal
elements $w$ are
naturally frozen into their ground states. This is correct but it is not quite legitimate,
in general, to ignore the effects of the zero point motion of $w$. The total ground state energy
of the oscillators is of order
\begin{eqnarray}  \displaystyle{
N_1 N_2 \omega = \frac{N_1 N_2 R \: |X-Y|}{l_{11}^3}
}  \end{eqnarray}
and would give rise to a confining potential between gravitons. This would
be completely unacceptable. Here is where supersymmetry comes to the
rescue. The point is
that each boson fluctuation is accompanied by a fermionic partner whose
zero point energy
exactly cancels the bosonic term. This is an exact consequence of
supersymmetry.

There is a very close analogy with the Higgs effect. The $X$ correspond to
scalar fields in the
adjoint of $U(N)$. The background
\begin{eqnarray}  \displaystyle{
X = \pmatrix{
X_{c.m.} & 0 \cr
0 & Y_{c.m.} \cr
}
}  \end{eqnarray}
breaks the $U(N)$ symmetry to $U(N_1) \times U(N_2)$ and the off diagonal
elements
are analogous to W-bosons which become massive as a consequence of the symmetry
breaking. The vanishing of the zero point energy is identical to the
vanishing of the ground
state energy density in super-Higgs theories.

Let us next consider the velocity dependence of the forces between supergravitons
 \cite{14} .
The simplest way to think about these interactions is to continue the analogy with the Higgs
effect. The effective action for slowly varying Higgs field can be computed by integrating out the
massive W fields. For example, the effective action for $X_{c.m.} -Y_{c.m.} \equiv r$ can be
computed to quadratic order in the velocities by computing Feynman diagrams with the external
lines representing the fields $\dot{X}_{c.m.}$ and $\dot{Y}_{c.m.}$. The diagrams are completely
analogous to vacuum polarization diagrams which renormalize the kinetic terms in QFT.
Fortunately the calculation is not necessary. The high degree of
supersymmetry insures that
the result exactly vanishes.

The first non-vanishing interaction arises at the quartic level in
velocities. It has been exactly
computed to the one and two loop order of the matrix quantum mechanics (0+1 field theory). Here we will only sketch the 
method. 

From the matrix theory lagrangian
we can read off the Feynman rules for the theory. For example the propagator for the massive $w$ is
\begin{eqnarray}  \displaystyle{
\Delta_F = \frac{R}{\omega^2 + m^2 }
}  \end{eqnarray}
\begin{eqnarray}  \displaystyle{
m \sim \frac{Rr}{l_{11}^3}
}  \end{eqnarray}
and the quartic bosonic vertex is
\begin{eqnarray}  \displaystyle{
{\mathrm vertex} = \frac{R}{l_{11}^6}
}  \end{eqnarray}

Consider the one-loop diagram with four external $X$ lines at zero frequency (Figure 1). It has the form
\begin{eqnarray}  \displaystyle{
\int \frac{d \omega}{(\omega^2 + m^2)^2} \frac{R^4}{(l_{11})^{12}} N_1 N_2
}  \end{eqnarray}
The factor $N_1 N_2$ is just because there are $N_1 N_2$ off diagonal
W-bosons to integrate
out. Such diagrams each contribute to the effective potential to order
$X^4$, but, as we have
argued, they all add up to zero as a consequence of supersymmetry. However
we are not
interested in the static potential, but rather in terms involving velocities. One finds that the high degree of supersymmetry implies that the first nonvanishing contributions are at the level 
of 4 powers of velocity. Roughly
speaking, this introduces two more propagators, so that the effective action will have the form
\begin{eqnarray} \label{eq90}
\displaystyle{
S_{eff} \sim |\dot{r}|^4 \int \frac{d\omega}{(\omega^2 + m^2)^4} \frac{N_1
N_2 R^4}{l_{11}^2}
\sim \frac{1}{m^7} \dot{r}^4 \frac{N_1 N_2 R^4}{l_{11}^3}
}  \end{eqnarray}
Now, using
\begin{eqnarray}  \displaystyle{
m= \frac{rR}{l_{11}^3}
}  \end{eqnarray}
we find
\begin{eqnarray}  \displaystyle{
S_{eff} \sim \left[  \frac{l_{11}^9}{r^7 R^3} N_1 N_2 \right] |\dot{r}|^4
= \frac{N_1 N_2 G_N}{r^7 R^3} |\dot{X} - \dot{Y}|^4
}  \end{eqnarray}
where $G_N$ is the 11-dim.~Newton constant.

The point of this effective action is that it can be used to compute the
scattering of two
supergravitons for large impact parameter. The simplest way to do this is
to convert
$S_{eff}$ to an interaction Hamiltonian. From the free term in the action
\begin{eqnarray}  \displaystyle{
S_{free} = \frac{N_1}{R} \frac{\dot{X}^2}{2} + \frac{N_2}{R}
\frac{\dot{Y}^2}{2}
}  \end{eqnarray}
we identify the velocities in terms of 9-dimensional momenta
\begin{eqnarray}  \displaystyle{
\dot{X} = \frac{P_1}{N_1} R
}  \end{eqnarray}
\begin{eqnarray}  \displaystyle{
\dot{Y} = \frac{P_2}{N_2} R
}  \end{eqnarray}
\begin{eqnarray}  \displaystyle{
|\dot{X} - \dot{Y}|^4 = ( \frac{P_1}{N_1} -  \frac{P_2}{N_2} )^4 R^4
}  \end{eqnarray}

The interaction Hamiltonian is obtained from Legendre transforming the
Lagrangian.
It has the form
\begin{eqnarray}  \displaystyle{
H_{eff} = R \left[ \frac{P_1^2}{2N} + \frac{P_2^2}{2N} +
\frac{N_1 N_2 G_N}{r^7} \left( \frac{P_1}{N_1} - \frac{P_2}{N_2}
 \right)^4 \right]
}  \end{eqnarray}
or, using $\displaystyle{P_- = \frac{N}{R}}$,
\begin{eqnarray}  \label{hoppe} \displaystyle{
H_{eff} = R \left[ \frac{P_1^2}{2P_-^{(1)}} +
\frac{P_2^2}{2P_-^{(2)}} + \frac{a}{R^2}
\frac{P_-^{(1)} P_-^{(2)} G_N}{r^7}
\left( \frac{P_1}{P_-^{(1)}} - \frac{P_2}{P_-^{(2)}}
 \right)^4 \right]
}  \end{eqnarray}
where $a$ is a numerical constant which has been computed
\cite{14}.

The interaction term can now be treated in Born approximation
giving rise to a scattering amplitude at large impact parameter.
The resulting scattering amplitude can be compared with a corresponding
scattering
amplitude obtained from tree
diagrams in supergravity. In order to make this comparison, we
must restrict the process to the kinematic situation in which the
momentum transfer is purely transverse.

Processes in which $P_-$ is exchanged correspond in matrix
theory to events in which D0-branes are transfered from one
cluster ($N_1$ branes) to the other ($N_2$). The process computed in this
section does not
involve exchange of $N$ and so it must be compared with supergravity
amplitudes with
no $P_-$ exchange. With this restriction, the supergravity
scattering amplitude arising from single graviton scattering
is found to exactly agree with the result obtained from
(\ref{hoppe}).

Originally, it was thought that this should generalize to a great
number of processes that
\begin{itemize}
\item{involve such large impact parameters that only tree
diagrams should contribute;}
\item{involve no exchange of $P_-$.}
\end{itemize}

For example, one might attempt a similar calculation for
three body scattering in which all three supergravitons are
far from another in tranverse space. The supergravity
calculation involves the diagrams of the form shown in
figure 2.   

If we allow all distances between gravitons to be proportional to
$r$ and velocities to $v$ the amplitude scales like
\begin{eqnarray}  \displaystyle{
G_N \frac{v^6}{r^{14}}
}  \end{eqnarray}

It is not difficult in matrix theory to find a contribution to the
effective action which has this form. It arises from a two loop
matrix theory diagram. However when calculated in detail the
supergravity and matrix theory computations disagree
 \cite{16}.

Although this disagreement is disappointing, it does not
necessarily threaten the validity of matrix theory.
Let us first consider the situation from the supergravity side. According to sect.~1.2, DLCQ it may be thought
of as compactification on a vanishingly small spacelike circle
with the longitudinal momentum kept fixed in units of
$\frac{1}{R_C}$. As we discussed in lecture 1, in QFT there are
zero modes which carry $\tilde{P}_- =0$ which must be
integrated out. In the limit $R_C \rightarrow 0$ these zero
modes become strongly coupling with coupling of order $\frac{1}{R_C}$.
There is no obvious reason why the diagrams
involving non linear processes in the zero mode sector should
adequately represent the correct physics \cite{11}.

In the present situation, we are considering the compactification
of 11-D sugra down to 10-D with a Newton constant of order
\begin{eqnarray}  \displaystyle{
G_{10} \sim \frac{l_{11}^9}{R_C}
}  \end{eqnarray}
Now, for given $R_C$, a length scale exists which defines the
region for which perturbation theory in $G_{10}$ is valid.
Thus for distance satisfying
\begin{eqnarray}  \displaystyle{
r > G_{10}^{1/8} = \frac{l_{11}^{9/8}}{R_C^{1/8}}
}  \end{eqnarray}
the tree approximation for the zero modes should be adequate.
However, the length scales which interest us in M-theory are
fixed in units of $l_{11}$. Thus as $R_C \rightarrow 0$
the tree approximation of sugra tells us nothing about the DLCQ
of M-theory. Possible exceptions to this would be special
amplitudes which supersymmetry protects. Thus we might conjecture that the
two body amplitude in order $\dot{X}^4$
is protected but that the three body amplitude in order
$\dot{X}^6$ is not. In fact there is evidence that this is the case.

Now consider the matrix theory side. In principle,  the amplitudes
for both two and three body scattering are subject to matrix
theory loop corrections. The form of these corrections is illuminating.
 As an example consider the loop corrections to
the $\frac{\dot{X}^4}{r^7}$ two body force. Simple power
counting shows that the leading corrections are a power series
in $\frac{l_{11}^3 N}{r^3}$, assuming $N_1$ and $N_2 \, \sim N$.

Thus, the one loop amplitude could be corrected by a function
\begin{eqnarray} \label{eq_number} \displaystyle{
{ F} \left( \frac{N l_{11}^3}{r^3}  \right)  = 1 + c_2
\frac{N l_{11}^3}{r^3} + c_3 \left( \frac{N l_{11}^3}{r^3}  \right)^2
+ \ldots
}  \end{eqnarray}
Now, for fixed $N$ as $r \rightarrow \infty$ the higher
corrections vanish. But this is not the limit that interests us.
Instead we should fix $\frac{r}{l_{11}}$ and let $N \rightarrow
\infty$. Therefore the correct physics is determined by the
behaviour of $F$ as its argument tends to infinity, not
to zero.

On the other hand we have seen that the one loop amplitude
exactly agrees with expected supergravity results. This suggests
that a supersymmetric nonrenormalization theorem may be at
work. Thus far no general theorem has been derived, but the
coefficient $C_2$ has been computed
 \cite{17}  and,
encouragingly, the result vanishes.

In the case of the three body amplitude studied in
 \cite{16} 
similar corrections to (\ref{eq_number}) are expected.
However in this case there is unlikely to be a nonrenormalization
theorem. The amplitude is almost certainly not constrained by
supersymmetry since we already know that two different
supersymmetric theories give different results, namely tree
diagram supergravity and one loop matrix theory. Thus, we believe that the correct conjecture is not that matrix
theory (at finite $N$) and tree diagram sugra should agree,
but rather
\begin{enumerate}
\item{when they agree, a nonrenormalization theorem will be
discovered; }
\item{when they do not, higher loop diagrams will not vanish. }
\end{enumerate}
This could be confirmed by a three loop computation of the
$\dot{X}^6$ effective action.


\section{Lecture 3 (Objects and dualities in Matrix theory)}
\subsection{2-branes and 5-branes}
In the last lecture we demonstated how the
Fock space of supergravitons arises out of matrix degrees of freedom.
M-theory has a wide variety of other objects, including membranes,
5-branes and black holes. In addition when the theory is compactified
it should have strings and D-branes. In the first half of this lecture
we will develop the theory of membranes and describe what is known
about 5-branes. Then we will move on to the theory of
compactification. This will allow us to explore the origins of
string theory dualities in Matrix Theory.

There are various membrane configurations that can occur in
uncompactified M-theory. Large but finite membranes with spherical,
toroidal or higher genus topology are possible. These membranes are
unstable,  as they will eventually collapse into a black hole. These
computations are of course  very complicated, even classically, and
have never actually been  carried out, but it's clear what's going to
happen: the ``surface  tension'' of the membrane is a force which
acts always in the same direction,  so the brane is going to shrink
till it approaches its Schwartzschild radius. We
can however take  very big (but not infinite) membranes for which the
decay time is  very long and which are so large that a semiclassical
description  is appropriate. We may also consider stable infinite
membranes which extend over an entire plane in the 10 dimensional
space. Finally, if the theory is compactified, there are stable finite
membranes wrapped on two cycles of the compactification manifold. In
this lecture we consider only the first two situations.


A membrane is endowed with a world-volume in the same
sense in which a string has a world sheet. The world volume is
a three dimensional space consisting of time and a two dimensional
space  locally parametrized by two coordinates. In this lecture we
will consider only the toroidal topology although generalization
to other topologies is not difficult.

The geometry of the world space of the membrane in Matrix Theory is
not conventional. It is a kind of space that mathematicians call a
non-commuting geometry
\cite{Connes}. Such geometries  generalize classical geometry
much the same way as quantum mechanics generalizes the phase
space of classical mechanics. Begin by defining  two angular
coordinates parametrizing the world space we call
$p,q$. However, the coordinates are not ordinary
variables. They are noncommuting with commutation relations

\begin{eqnarray} \label{Number} \displaystyle{
[p, q] = \frac{2 \pi i}{N}
}  \end{eqnarray}

The non-commuting torus geometry allows us to construct 
a useful representation for all $N \times N $ matrices.
To this aim,
a more convenient labeling is
\begin{eqnarray}  \displaystyle{
U= e^{ip}
}  \end{eqnarray}
\begin{eqnarray}  \displaystyle{
V= e^{iq}
}  \end{eqnarray}
so that
\begin{eqnarray}  \displaystyle{
UV = e^{\frac{2 \pi i}{N}} VU
}  \end{eqnarray}
We can now represent any $N \times N$ matrix in
terms of the $U$, $V$, just by slightly generalizing the Fourier
sum expansion:
\begin{eqnarray} \label{EQUATIOn109} \displaystyle{
Z = \sum_{n,m=1}^{N}{Z_{nm} (U)^n (V)^m}
}  \end{eqnarray}

The representation (\ref{EQUATIOn109}) defines a function of two periodic
variables for every $N \times N$ matrix.
As long as $N$ is finite we have  operator ordering
ambiguities, but when $N$ is very large  the
description becomes semiclassical. In fact from 
(\ref{Number}) we see
that $1/N$ plays the role of Planck's  constant and we  encounter
the  same pattern of the
$\hbar
\rightarrow 0$ limit  as in ordinary quantum mechanics. In these
lectures we will not consider the issue of convergence to the
classical limit. However the reader should be aware that the
convergence is subtle and not all configurations have a good classical
description even as $N \to \infty$.

The mathematical correspondence between matrices and functions of
$p,q$ includes rules for replacing the trace operation and the
commutator.

\begin{eqnarray}  \displaystyle{
Tr \: Z  \rightarrow N \int dp \: dq Z(p,q)
}  \end{eqnarray}
\begin{eqnarray}  \displaystyle{
[Z, W] \rightarrow \frac{i}{N} \left[
{\partial{Z} \over \partial q}   {\partial{W} \over \partial p}
- {\partial {Z}\over \partial p}  {\partial {W}\over \partial q}
\right]  = \frac{i}{N}  \{  Z, W  \}_{\mathrm Poisson \: bracket}
}  \end{eqnarray}

Using these semi-classical rules we can now rewrite the matrix
Lagrangian
\begin{eqnarray}  \displaystyle{
{\cal L} = Tr \left\{ \frac{1}{2R} \dot{X}^i \dot{X}^i + 
\frac{R}{4 \: l_{11}^6}
[X^i, X^j]^2 + {\mathrm fermions}   \right\}
}  \end{eqnarray}
in the large $N$ limit, inserting the above replacements:
\begin{eqnarray}  \label{3.7} \displaystyle{
H = \frac{R}{N} \int dp \: dq \left\{ \frac{1}{2}
\left( \Pi_i (p,q)  \right)^2
- \frac{1}{4 \: l_{11}^6} \{  X^i, X^j  \}^2_{\mathrm Poisson \: bracket}
+ {\mathrm fermions}   \right\}
}  \end{eqnarray}

This Hamiltonian obviously describes a toroidal membrane whose
embedding in transverse space is given by $X(p,q)$. In fact it
is precisely the Hamiltonian of a supermembrane in light cone frame \cite{21}. 


The first thing to note about eq.~(\ref{3.7}) is the proportionality
to $\displaystyle{\frac{R}{N}= \frac{1}{P_-}}$; the energy
excitations go as $\displaystyle{\frac{1}{P_-}}$, which is the right
kind of behaviour for a localized object in the light cone frame. In this way we  have  tested, even if  only semiclassically,
something which was not obvious; from  now on, we can treat the $p$
and $q$ as ``true'' coordinates  and assume we have recovered the
previously studied two-brane.

 We now turn to  the infinite 2-brane. The rigorous way to do this is
to first compactify  the theory on a 2-torus of size $L \times L$ and
consider a membrane wrapped around the torus. The  limit $L \to
\infty$ defines the infinite membrane. However, since we have not yet
studied compactification we will adopt an informal less rigorous
approach.

To this aim let's
introduce an IR cutoff $L$.  We  then define
\begin{eqnarray}  \displaystyle{
X^1 = \frac{pL}{2 \pi}
}  \end{eqnarray}
\begin{eqnarray}  \displaystyle{
X^2 = \frac{qL}{2 \pi}
}  \end{eqnarray}

The area of the membrane in classical physics would be obtained from
the integral of the Poisson bracket of $X^1, X^2$. The correspondence
with matrices gives

\begin{eqnarray}  \displaystyle{
\frac{1}{N} Tr [X^1, X^2] \sim L^2 \equiv
{\mathrm Area \: of \: the \: I.R. \:(cut \: off) \: 2-brane}
}  \end{eqnarray}

\par Let us now turn to another example \cite{19}. This time we 
are going to
consider a new kind of object. So far the objects we have studied can
be thought of as longitudinally localized "lumps" which look 
like blurred world lines. These objects Lorentz
contract as $N$ increases. They are characterized by having energy
which decreases like ${1 \over P_-} = {R\over N}$. Now we are going
to consider a brane which is wrapped around the light like direction
$X^-$. Consider a  p-brane with tension $T$ with one of its
dimensions wrapped around $X^-$ and the other $p-1$ dimensions
stretched out over a volume $L^{p-1}$. The energy of such an object is $E=TL^{p-1}R$. 
Note that it scales like $R$ but does not contain the factor $1/N$.

Consider the following configuration.
Let $N=n^2$. Define $p$ and $q$ to be $n \times n$ matrices 
which satisfy $\displaystyle{[q,p]= \frac{2 \pi i}{n}}$. Let
\begin{eqnarray}  \displaystyle{
X^1= p \otimes I \cdot L
}  \end{eqnarray}
\begin{eqnarray}  \displaystyle{
X^2= q \otimes I \cdot L
}  \end{eqnarray}
\begin{eqnarray}  \displaystyle{
X^3= I \otimes p \cdot L
}  \end{eqnarray}
\begin{eqnarray}  \displaystyle{
X^4= I \otimes q \cdot L
}  \end{eqnarray}
The matrix Hamiltonian for such static configurations becomes
\begin{eqnarray}  \displaystyle{
E= \frac{R}{4 \: l_{11}^6}  
\: Tr \left\{ [X^i, X^j]^2 \right\} =  \frac{R L^4}{4 \: l_{11}^6}  \: \frac{1}{n^2} \:
Tr (I \otimes I) = \frac{R L^4}{4 \: l_{11}^6} 
}  \end{eqnarray}
That is
\begin{eqnarray} \label{3.16} \displaystyle{
E \sim \frac{L^4 R}{l_{11}^6}
}  \end{eqnarray}
Notice that energy goes as $R$, not as $\displaystyle{ 
\frac{R}{N}}$ as for
``lumplike objects''. Thus the object is a wrapped brane. Evidently
it is a wrapped 5-brane of some sort since its energy scales like 5 powers of a length. The 5-brane discussed in this section is not a "pure" 5-brane. It also has 2-brane charge \cite{19}. 
It is a 5-brane with
"dissolved" 2-branes oriented in the 1,2 and the 3,4 planes. However
it is the simplest 5-brane configuration to describe in terms of
Matrix Theory.

Presumably the theory also contains various kinds of 5-branes not
wrapped on $X^-$ but these are more difficult to describe. 

\subsection{Compactification}
\par To go beyond, and in particular to test dualities, we have to
compactify the theory. To avoid possible confusion, we want to
point out that up to now we have a theory with a strange
(lightlike) compactification along $X^-$, that is, in the
longitudinal plane. We will undo this in the end, but in any case
the $X^-$  compactification has nothing to do with true
(spacelike) compactification. This we will choose to do in the
transverse space. 
After such compactification of a single direction matrix theory 
becomes IIA string theory in light cone frame and we have all of 
the KK modes discussion of par.~2.1.

So, let's proceed to compactify along a (matrix!) direction.
Usually the compactification is carried out along the 11th
direction; we are going, instead, to do it along $X^9$. 

There is a recipe for compactifying a matrix direction \cite{22}.
Consider infinite
matrices of the form (where $X^i(n)$ are $N \times N$ matrices)
\begin{eqnarray}  \displaystyle{
X^i=
\left (
\begin{array}{ccccc}
X^i (1) & X^i(2) & X^i(3) & \cdots & \cdots \\
\cdots   & X^i(1) & X^i(2) & X^i(3)& \ddots \\
\vdots & . & \ddots &  X^i(2) & \ddots \\
. & .         & \cdots        & X^i(1)  & \ddots \\
\end{array}
\right )
\: \: \: \: \: \: \: \: \: \: \: \: \: i \neq 9
}  \end{eqnarray}
\begin{eqnarray}  \displaystyle{
X^9=
\left (
\begin{array}{cccc}
X^9 (1)  - 2L & X^9(2) & X^9(3) & \cdots \\
\cdots   & X^9(1) - L & X^9(2) & \ddots \\
\vdots & . & X^9(1) &  \ddots \\
. & .    & .      & X^9(1) +L     \\
. & .         & \cdots        & \ddots
\end{array}
\right )
}  \end{eqnarray}
that is, the directions are $N \times N$ block matrices with suitable
shifts of $L$, $2L$, etc.~on
the diagonal for the compact direction. (This is the usual notion of
compactification: we can
imagine to put side by side various copies of the compactified space, which
are identical but
shifted by the compactification length $L$).

We can write ${\cal L}$ in terms of suitable ``matrix fields''.
We have a matrix representation

\begin{eqnarray} \displaystyle{
\left (\matrix{
-2L & 0 & 0 & 0 & 0\cr
0 & L & 0 & 0  & 0 \cr
0 & 0 & 2L & 0 & 0 \cr
0 & 0 & 0 & 3L & 0 \cr
0 & 0 & 0 & 0 & \ddots \cr
}\right )
= - i L  {\partial \over \partial \sigma}
} \end{eqnarray}

\begin{eqnarray}  \displaystyle{
\left (\matrix{
0 & 1 & 0 & 0 & \cdots \cr
0 & 0 & 1 & 0 & \cdots \cr
0 & 0 & 0 & 1 & \cdots \cr
0 & 0 & 0 & 0 & \cdots \cr
\cdots & \cdots & \cdots & \cdots & \cdots \cr
1 & 0 & 0 & 0 & \cdots \cr
}\right )
= e^{i \sigma}
}  \end{eqnarray}

\begin{eqnarray}  \displaystyle{
\left (\matrix{
0 & \cdots & 1 & 0 & \cdots \cr
0 & 0 & \cdots & 1  & \cdots \cr
0 & 0 & 0 & \cdots & \cdots \cr
\cdots & \cdots & \cdots & \cdots  & \cdots \cr
}\right )
= e^{in \sigma}
}  \end{eqnarray}
\begin{eqnarray}  \displaystyle{
X^9 = i {\partial \over \partial \sigma} + \sum{X^9 (n) e^{i n \sigma}} = i
{\partial \over
\partial \sigma} + A(\sigma)
}  \end{eqnarray}
\begin{eqnarray}  \displaystyle{
X^i = \phi^i(\sigma) \: \: \: \: \: \: \: \: \: \: \: \: \: i \neq 9
}  \end{eqnarray}
\par Let us work out the terms of the Lagrangian:
\begin{eqnarray}  \displaystyle{
{\mathrm Kinetic \: terms} \Longrightarrow \frac{1}{R} \int d \sigma
\left( \frac{\dot{A}^2}{2} + \frac{\dot{\phi}^2}{2}  \right)
}  \end{eqnarray}
\begin{eqnarray}  \displaystyle{
[X^i, X^j]^2 \: , \: \: \: \: i \neq j \neq 9 \Longrightarrow R \int d
\sigma [\phi^i(\sigma,
\phi^j(\sigma)]
}  \end{eqnarray}
\begin{eqnarray}  \displaystyle{
[X^i, X^9]^2 \Longrightarrow R \int d\sigma \left( {\partial {\phi^i}\over
\partial \sigma}
+ i [A, \phi^i] \right)^2 = R \int d \sigma |D_\sigma \phi |^2
}  \end{eqnarray}

Similar considerations apply to fermions.

This is a super Yang Mills theory; in this case, it is endowed with 16 susy
and lives in 1+1
dimensions, over a (dual) 1-torus. After suitable rescalings of fields and of
$\sigma$ the Lagrangian appears as ($\phi$ is the $X$ field after rescaling)
\begin{eqnarray}  \displaystyle{
{\cal L} = \frac{1}{g_{YM}^2} \int_0^\Sigma \left( \frac{\dot{A}^2}{2} +
\frac{\dot{\phi}^2}{2}
- \frac{(D \phi)^2}{2} + {\mathrm fermions}
\right)
}  \end{eqnarray}
where $\Sigma$ is the compactified radius of the field theory and $g_{YM}$
its coupling
constant. We have thus three quantities, $L$, $\Sigma$ and $g_{YM}$, of
which only $L$ was
known before; actually, the theory before had really just one parameter. Thus
we will have to relate to $L$ the other two.
\par Let us, first of all, summarize our notations. The rules on a
$d$-torus are basically
the same. M theory lives on a torus (chosen to be
transverse) of sides $L_i$ and volume $V= \prod_{i}{L_i}$; it has only one
physical
parameter, in addition to $L_i$: the eleven dimension Planck scale $l_{11}$.
However, it contains another (unphysical)
length scale, the radius of lightlike compactification $R$. The super Yang
Mills theory lives
on a dual torus of sides $\Sigma_i$ and volume $V_D = \prod_i \Sigma_i$. To
describe the
YM parameters in terms of $L_i$ and $l_{11}$, let's remember again that $\dot{\phi}$ and
$\dot{A}$ are the velocities (derivatives of $X$) respectively in the
noncompactified and
compactified direction, and accordingly the electric energy is the kinetic
energy for the
compact directions. The electric term of YM theory is
\begin{eqnarray}  \displaystyle{
L \sim \frac{1}{g_{YM}^2} \int d^dx \frac{\dot{A}_i^2}{2} \sim
\frac{1}{g_{YM}^2}  V_D
\frac{\dot{A}_i^2}{2}
}  \end{eqnarray}

\begin{eqnarray}  \label{3.29} \displaystyle{
H= \frac{g_{YM}^2}{2 V_D} \Pi^2_{A_i} = \frac{g_{YM}^2}{2 V_D} \Sigma_i^2 n^2
}  \end{eqnarray}
where
\begin{eqnarray}  \displaystyle{
\Pi_{A_i} = {\partial {\cal L}\over \partial {\dot{A}} }
}  \end{eqnarray}

Eq.~(\ref{3.29}) and in particular the interpretation of $A_i \Sigma_i$ as
an angle and
the quantization rule $\Pi_{A_i}= \sum_i n$ arise from the Wilson loop
behaviour of a
electromagnetic-like theory on a circle, that is, after one direction has
been compactified.
In principle one needs the theory to be abelian, but if this is not the
case it is enough to
refer to the $U(1)$ contained in the gauge group. The $U(1)$
we are referring to corresponds to separating off one
zero-brane.

M theory gives for the kinetic energy the expressions
\begin{eqnarray}  \label{3.30} \displaystyle{
{\mathrm kinetic \: energy} = \frac{p_i^2}{2 p_-} = \frac{n^2 R}{2 L_i^2}
}  \end{eqnarray}
where $\displaystyle{P_-=\frac{1}{R}}$, 
$\displaystyle{P_i = \frac{n}{L_i}}$.

Comparison between (\ref{3.29}) and (\ref{3.30}) gives
\begin{eqnarray}  \label{3.31} \displaystyle{
g^2_{YM} = \frac{V_D R }{L_i^2 \Sigma_i^2}
}  \end{eqnarray}
Another equation for $\Sigma$ can be obtained.
We remember that in the SYM picture
a KK quantum of momentum $N/ \sigma_i$ has energy
\begin{eqnarray}  \displaystyle{
E = \frac{n}{\Sigma_i}
}  \end{eqnarray}
The correspondent matrix theory energy is the one of the winding modes
\begin{eqnarray}  \displaystyle{
E = \frac{n L_i R}{l_{11}^3}
}  \end{eqnarray}
Equating these we get
\begin{eqnarray}  \label{3.34} \displaystyle{
L_i \Sigma_i = l_{11}^3 \frac{1}{R}
}  \end{eqnarray}
and
\begin{eqnarray}  \label{3.35} \displaystyle{
V_D = \frac{l_{11}^{3d}}{R^d} \frac{1}{V}
}  \end{eqnarray}
Inserting (\ref{3.34}) and (\ref{3.35}) in (\ref{3.31}) gives
\begin{eqnarray} \label{141} \displaystyle{
g_{YM}^2 = \frac{l_{11}^{(3d-6)} R^{3-d}}{V}
}  \end{eqnarray}
We can also introduce a dimensionless coupling
\begin{eqnarray}  \displaystyle{
\tilde{g}^2 = \frac{l_{11}^3}{L^3} \equiv g_{YM}^2 \Sigma^{d-3}
}  \end{eqnarray}
whose expression does not depend on $d$.

\subsection{T-duality}
We would now like to see that matrix theory embodies the expected
dualities of M theory and string theory \cite{23}, \cite{24}, \cite{25}. 
Let us act with T-duality on the strings. We know that the result is
\begin{eqnarray}  \displaystyle{
L_i \longrightarrow \frac{l_s^2}{L_i}  \equiv \tilde{L}_i
 }  \end{eqnarray}
for all the $L_i$ over which we dualize (a subset of the total). We know
also that
T-dualizing over an odd number of $L_i$ interchanges IIA and IIB string
theories,
while doing it over an even number of $L_i$ sends them respectively in
themselves.

To study the even case we must have at least $L_i$, $i$=1, 2, 3 (which
means a string theory
living in 8 dimensions); we choose $L_1$ to be a ``dilaton direction'' and
will dualize
$L_2$, $L_3$. We have
\begin{eqnarray}  \displaystyle{
\tilde{L}_2 = \frac{l_s^2}{L_2}
}  \end{eqnarray}
\begin{eqnarray}  \displaystyle{
\tilde{L}_3 = \frac{l_s^2}{L_3}
}  \end{eqnarray}
and
\begin{eqnarray}  \displaystyle{
\tilde{l}_s = l_s
}  \end{eqnarray}
\begin{eqnarray}  \displaystyle{
\tilde{G}_N^{(8)} = G_N^{(8)}
}  \end{eqnarray}
One has also
\begin{eqnarray}  \displaystyle{
l_s^2 = \frac{l_{11}^3}{L_1}
}  \end{eqnarray}
where $l_s$ is the characteristic string length if we think of a string as
limit of
a wrapped 2-brane:
\begin{eqnarray}  \displaystyle{
l_s^2 = \frac{l_{11}^3}{R}
}  \end{eqnarray}

The effect of T-duality is thus, explicitly in terms of $L_i$, $i=1, ... 3$
and $l_{11}$,
\begin{eqnarray}  \displaystyle{
\tilde{L}_1 = \frac{l_{11}^3}{L_2 L_3}
}  \end{eqnarray}
\begin{eqnarray}  \displaystyle{
\tilde{L}_2 = \frac{l_{11}^3}{L_2 L_1}
}  \end{eqnarray}
\begin{eqnarray}  \displaystyle{
\tilde{L}_3 = \frac{l_{11}^3}{L_1 L_3}
}  \end{eqnarray}
\begin{eqnarray}  \displaystyle{
\tilde{l}_{11}^3 = \frac{l_{11}^6}{L_1 L_2 L_3}
}  \end{eqnarray}

Using (\ref{141}) for the dual torus, it turns
out that
\begin{eqnarray}  \displaystyle{
\tilde{g}_{YM}^2 {g}_{YM}^2  = (2 \pi)^2 
}  \end{eqnarray}
which has the form of electric/magnetic duality. This SYM theory is actually believed to
enjoy exact Montonen-Olive duality. It should be noted that the 
argument presented in this lecture is the first demonstration 
of T-duality which does not depend on ordinary string 
perturbation theory. 

We will come back to this topic in the next
section where we will discuss how a new (and unexpected) spatial direction
can arise in the
theory if we shrink a 3-torus to a 2-torus and, more generally, what is the
relation between
coordinates and fluxes (which gives to space a very special and unusual
role). For the moment,
let us summarize that (by means of the identification between electric and
magnetic fields and
velocities in noncompact and compact directions) the electric/magnetic
duality, which
exchanges quanta of electric flux and of magnetic flux, is translated, in
the matrix theory
picture,  as a T-duality which interchanges momentum in the compact
direction with the quanta
(corresponding to topological quantum numbers) representing winding of
strings or wrapping
of membranes.

\section{Lecture 4 (The emergence of space in Matrix theory)}
The role of space in matrix theory is very unusual if compared to what happens in a field theory. Spatial directions materialize in several new and surprising ways, some of which we will now list. 
\begin{itemize}
\item{The usual transverse coordinates $X$ arise as the moduli 
space of a super Yang Mills theory. }
\item{The longitudinal direction $X^-$ is described as the conjugate to $P_-$ which is proportional to the $N$ of $U(N)$ 
super YM theory. }
\item{New unexpected directions can emerge as in the case of 
compactification on a 2-torus. As the 2-torus shrinks to zero a 
new 10th direction opens up; it is connected to the Yang Mills 
magnetic flux.} 
\end{itemize}

Our guide to understanding of the above properties will be the symmetries,
and in particular rotation invariance; the unexpected features will,
in the following, appear as unexpected extensions of these symmetries.

We have discussed, in the previous chapter, the relation between
electric/magnetic
duality in the super Yang-Mills gauge theory and T-duality on the matrix theory side. 
In that case we considered a
theory on the $T^3$ torus; now we move to a 2-torus. For 
simplicity we work only with a rectangular torus of sides $L_1$, 
$L_2$. Now consider the limit in which the face of the torus 
is ``shrunk'': 
\begin{eqnarray}  \displaystyle{
L_1 L_2 \rightarrow 0 \: \: , \:\:\:\:\:\:\:\:\:\: \frac{L_1}{L_2} \:
{\mathrm fixed}
}  \end{eqnarray}
This causes on the YM side the blowing up of the dual torus. As $L_1 L_2$ is shrunk, the cost of a unit of magnetic flux goes down (the magnetic
field is inversely proportional to the area and the energy stored in it goes as $B^2 \cdot$ area).  The energy turns out to be 
proportional to the area and to $l_{11}^{-3}$:
\begin{eqnarray} \label{156} \displaystyle{
E= \frac{L_1 L_2}{l_{11}^3}
}  \end{eqnarray} 
It is evident that this configuration describes a membrane of 
tension $1/{l_{11}^3}$ wrapped on the 2-torus. The energy cost 
decreases, leading to an accumulation of a vast number of very low energy states. As we will see, these states can be 
identified with a new spatial dimension, $Y$, decompactifying 
as $L_1 L_2 \rightarrow 0$ \cite{6}.  The  new direction is identified through its conjugate momentum which is obviously proportional to the two brane wrapping number or equivalently the yang mills magnetic flux which we call $n$. 

Let's compute the 
compactification radius $L_Y$ of the hypothetical new direction. 
The energy of a KK mode along the $Y$ direction is
\begin{eqnarray} \label{157} \displaystyle{
E = \frac{1}{L_Y}
}  \end{eqnarray}
Equating (\ref{156}) and (\ref{157}) gives the 
volume of the 3-torus ($Y$,1,2):
\begin{eqnarray}  \displaystyle{
L_Y L_1 L_2 = l_{11}^3
}  \end{eqnarray}

For $L_1$, $L_2$ finite $L_Y$ is also finite, but when $L_1 L_2 \rightarrow 0$,
$L_Y \rightarrow \infty$ and the transverse space is generated by one more
vector:
it is the span of $\{  X^3, \ldots, X^9, Y \}$. 

How do we know that the correct interpretation of the low 
energy states is in terms of a new direction of space? It would 
obviously be sufficient to prove the existence of a symmetry which rotates $Y$ into one of the noncompact directions 
$(X^3, \ldots, X^9)$. In order to demonstrate such an invariance 
we will temporarily compactify $X^3$ so that its radius is 
identical to $L_Y$. The theory now becomes a 3+1 SYM theory 
which was discussed in the previous lecture. The choice $L_3= 
L_Y$ gives \cite{26} 
\begin{eqnarray}  \displaystyle{
L_1 L_2 L_3 = l_{11}^3
}  \end{eqnarray}
and using (\ref{141}) we compute the Yang-Mills coupling
\begin{eqnarray}  \displaystyle{
g^2_{YM} = 2 \pi
}  \end{eqnarray}
Thus the YM theory is at its self dual point at which it enjoys the 
symmetry
\begin{eqnarray}  \displaystyle{
E_i \rightarrow B_{ij} \epsilon_{ijk}
}  \end{eqnarray}

Now the electric field $E_i$ is the conjugate to $A_i$, which also means that it is momentum conjugate to $X^3$. Using 
\begin{eqnarray}  \displaystyle{
P_3 = E_3
}  \end{eqnarray}
\begin{eqnarray}  \displaystyle{
P_Y = B_{12} \epsilon_{123}
}  \end{eqnarray}
we see that there is a discrete symmetry ($P_3 \leftrightarrow 
P_Y)$. 

We have then a set of spatial directions
\begin{eqnarray}  \displaystyle{
(Y; X^3, X^4, \ldots , X^9)
}  \end{eqnarray}
which enjoys the (discrete) exchange symmetry between $Y$ and $X^3$. On the other hand, in the limit $L_1 L_2 \rightarrow
0$, $L_3$ becomes infinite and the $O(7)$ invariance 
is restored. The only way the theory can have $O(7)$ invariance 
and $Y \leftrightarrow X^3$ invariance is for it to become fully 
$O(8)$ invariant. Thus, if we let $L_Y$ and $L_3$ go to infinity, 
we'll regain $O(8)$, the complete continuous rotational 
symmetry. 

There is another way to see the $O(8)$ invariance in the limit 
$L_1, L_2 \rightarrow 0$. Returning to the 2+1 dimensional 
theory, eq.~(\ref{141}) gives 
\begin{eqnarray}  \displaystyle{
g^2_{YM} = \frac{(l_{11})^{3d-6}}{V} R^{3-d}
}  \end{eqnarray}
and for its dimensionless analogue
\begin{eqnarray}  \label{143} \displaystyle{
\tilde{g}_{YM} = \frac{l_{11}^3}{(L_1 L_2)^{3/2}}
}  \end{eqnarray}
Since $\tilde{g}_{YM}$ is dimensionless it makes sense to 
discuss if it is large or small; let's notice, then, that it goes to
infinity as the torus shrinks. If we prefer we can state that, keeping fixed the dimensional coupling constant,
the YM torus becomes larger and larger. In the limit we are 
driven to a strongly coupled scale
invariant fixed point of a SCFT, whose behaviour has been classified. In the present case, it is known that the relevant 
fixed point has $O(8)$ invariance \cite{Seiberg}. 

\subsection{Supergraviton scattering with exchange of $P_Y$}
In this section we will discuss supergraviton scattering with 
exchange of momentum in
the $Y$ direction \cite{27} (but no exchange of $P_-$). 
Before doing that, we suppose that the two graviton scattering involves no
transfer either of
longitudinal or $Y$ momentum (for uncompactified theory in
11 dim.)
\begin{eqnarray} \label{165} \displaystyle{
\Delta P_Y = \Delta P_- = 0
}  \end{eqnarray}
Ti is not difficult to generalize the argument of section 2.4 
to the case of scattering with vanishing longitudinal and $Y$ 
momentum. The only difficulty is that the momentum integral 
in eq.~(\ref{eq90}) is now replaced by an integration over 
$d^3 k$, since the matrix theory is now a 2+1 dimensional 
quantum field theory. We obtain 
\begin{eqnarray}  \displaystyle{
\dot{X}^4 \int \frac{d^{3}k}{k^8} \sim \frac{\dot{X}^4}{\rho^{5}}
}  \end{eqnarray}
where $\rho$ is distance in the 7 dimensional space 
or impact parameter. This again agrees with the supergravity 
tree diagram analysis. 

%
%

We are now going to study processes where 
 $\Delta P_Y \neq 0$. We
are going to
argue that the  exchange of momentum is actually exchange of magnetic flux
and that the whole
process is an instantonic one
\cite{Polchiski_Pouliot}; it corresponds actually to the creation of a Polyakov instanton. 

Let us consider two supergravitons of longitudinal momentum
$\displaystyle{\frac{N}{2}}$ each:
\begin{eqnarray} \label{148} \displaystyle{
\left(
\begin{array}{cc}
(X^I) & \cdot \\
\cdot & (X^{II}) \\
\end{array}
\right)
}  \end{eqnarray}
$X^I$, $X^{II}$ are matrices $\displaystyle{\frac{N}{2} \times
 \frac{N}{2}}$; $X^I - X^{II}$ is very
large. The $U(N)$ group breaks to a $U(\frac{N}{2}) \times U(\frac{N}{2})$.

If we work in transverse center of mass coordinates, (\ref{148}) becomes
\begin{eqnarray}  \displaystyle{
\left(
\begin{array}{cc}
(X) & \cdot \\
\cdot & (-X) \\
\end{array}
\right)
}  \end{eqnarray}
Now, the two $U(1)$s in $U(\frac{N}{2}) \times U(\frac{N}{2})$ are the
center of
mass position of
the supergravitons and the two abelian magnetic fluxes 
$n_1$, $n_2$ are the $Y$ momenta.

Let us now consider a process in which a single unit of 
magnetic flux is exchanged: 
\begin{eqnarray}  \displaystyle{
n_1 \rightarrow n_1 +1
}  \end{eqnarray}
\begin{eqnarray}  \displaystyle{
n_2 \rightarrow n_2 -1
}  \end{eqnarray}
Another way to describe this process is that a unit of magnetic flux is created in a $U(1)$ contained in a $SU(2)$ subgroup of 
$U(N)$. Such a process in which an $SU(2)$ magnetic flux 
changes is actually an instanton process in 2+1 dimensions. 

Thus the process we are considering is an instanton in an 
$SU(2)$ theory spontaneously broken to $U(1)$. The Higgs 
expectation value is proportional to the separation of the 
gravitons with the precise connection given by 
\begin{eqnarray} \displaystyle{ 
\langle \phi \rangle = g^2_{YM} \rho / L_Y 
} \end{eqnarray}

The instanton action is
\begin{eqnarray}  \displaystyle{
S_{inst} = \frac{2 \pi}{g_{YM}^2} \langle \phi \rangle = 2 \pi \frac{\rho
L_1 L_2}{l_{11}^3}
= \frac{2 \pi \rho}{L_Y}
}  \end{eqnarray}
where $\langle \phi \rangle$ is the v.~e.~v.~ of the Higgs field
and, if we remember that the moduli space is the $X$ space,
is not only a v.~e.~v.~but also a distance; that is, $\rho$ is
to be interpreted as the distance between the branes.

The amplitude for the exchange of one unit of momentum in the $Y$ direction
in the regime $\rho \gg L_Y$ is proportional to
\begin{eqnarray}  \displaystyle{
\frac{\displaystyle{v^4 e^{- 2 \pi \rho / L_Y}}}{\rho^3}
}  \end{eqnarray}
The exponential suggests that this is the formula for an exchange process and this
is indeed the case: the mass for the Kaluza-Klein first mode is
\begin{eqnarray}  \displaystyle{
M_{KK} = \frac{2 \pi}{L_Y}
}  \end{eqnarray}
and the corresponding amplitude is straightforwardly proportional to
\begin{eqnarray}  \displaystyle{
e^{-(M_{KK} \rho)} = e^{- \frac{2 \pi \rho}{L_Y}}
}  \end{eqnarray}

Thus we see how exchange processes involving $Y$ momentum 
are generated by nonperturbative instanton processes 
\cite{BFSS2}.

In these lectures we have considered matrix theory compactified 
on 2 and 3-tori. The case on the 1-torus is also well studied. As we might expect, in the limit of small compactification radius the theory on a 1-torus becomes IIA string theory in the light cone frame. This has been beautifully demonstrated in the papers by Motl, Banks and Seiberg and Dijkgraaf, Verlinde and Verlinde. 
Unfortunately there is not enough time in this lectures to review 
their extremely interesting work. 

The theory compactified on $T^d$ with $d>3$ is much more 
problematic. In this case the relevant SYM theories are non 
renormalizable and therefore require new short distance data to 
define them. Interesting proposals for the theory on 4,5 and 6 
tori exist in the literature, but they are beyond the scope of this 
elementary review. 

Finally, the case of $T^d$ which would define a world with 3+1 
noncompact dimensions appears to require an entirely new 
approach.

\section{Lecture 5 (Black holes in matrix theory)} 
\subsection{Choosing $N$}
Our focus in these lectures has been on identifying and studying 
all the various objects that matrix theory needs to describe. 
We have seen how supergravitons, membranes and 5-branes 
arise. In addition, when compactified on a small $S^1$, weakly 
coupled IIA string theory emerges \cite{28}. This leaves only 
black holes. 

From the traditional relativists' point of view, black holes are 
extremely mysterious objects. They are described by unique 
classical solutions of Einstein's equations. All perturbations 
quickly die away leaving a featureless ``bald'' black hole with 
``no hair''. On the other hand Bekenstein and Hawking have 
given persuasive arguments that black holes possess 
thermodynamic entropy and temperature which point to the 
existence of a hidden microstructure. In particular, entropy 
generally represents the counting of hidden microstates which 
are invisible in a coarse grained description. In this lecture we 
will address the problem of black hole entropy in matrix theory. 

Most string theory studies of black hole entropy have 
concentrated on BPS black holes and their low lying excitations. 
Supersymmetry allows an extraordinary degree of mathematical 
control over these objects and quite precise state counting can be done. The results are in remarkable agreement with the 
Bekenstein Hawking formula. Much of this work can probably be 
translated into matrix theory terms \cite{30} 
but in this lecture we will concentrate on another class of black holes -the Schwartzschild 
black holes. These objects are as far as possible from BPS 
supersymmetric states and are therefore uncontrolled by the 
constraints of supersymmetry. We will concentrate on a 
particular special case, which in many respects is the simplest, 
Schwartzschild black holes in $7+1$ dimensions, obtained from 
matrix theory compactified on a 3-torus. 

An ultimate exact treatment of objects in matrix theory requires 
a passage to the infinite $N$ limit. Unfortunately this limit is 
extremely difficult. The limit is in many respect similar to taking 
the continuum limit of a lattice gauge theory. In the limit of 
vanishingly small lattice spacing, $a \rightarrow 0$, the number 
of degrees of freedom used to describe a given hadronic system diverges. Obviously there is no real need to describe a 
system by so many d.~o.~f.: a coarse grained description in 
which the lattice spacing $a$ is a few times smaller than the 
probed wavelengths is sufficient to capture the relevant 
properties. In this sense we may introduce the idea of a 
maximum lattice spacing $a_{max}$ which will allow an accurate 
computation of the properties of the system. The value of 
$a_{max}$ depends on the system or process. For example, if 
we are interested in inelastic electroproduction, the value of 
$a_{max}$ must decrease as the momentum transfer increases. 

Very similar remarks apply to the choice of $N$. The $N 
\rightarrow \infty$ limit introduces far more d.~o.~f.~than are 
actually needed to study any given system. Therefore our first 
task in applying matrix theory as a quantitative tool is to choose 
a value of $N$ which is large enough to capture the physics of a 
given black hole. Thus let us define $N_{min}(S)$ to be the 
minimum value of $N$ which will allow a given degree of 
accuracy in calculating properties of a black hole of entropy 
$S$. 

The value of $N_{min}$ may be obtained from a simple 
spacetime argument in lightlike compactification of M-theory. 
The basic requirement of an accurate description of a given 
system is that the compactification radius $R$ be large enough 
to easily contain the system without distortion. Thus consider a 
black hole of mass $M$ and Schwartzschild radius $R_s$. Let 
us choose the black hole to have momentum 
\begin{eqnarray}  \displaystyle{
P_+ = P_- \sim M
}  \end{eqnarray}
\begin{eqnarray}  \displaystyle{
P_i = 0 
}  \end{eqnarray}
In other words, the black hole is at rest. In this frame, the 
condition for the black hole to fit safely in the compact 
longitudinal space is 
\begin{eqnarray}  \displaystyle{
R > R_s
}  \end{eqnarray}
Now, multiplying both sides by $P_-$, we find 
\begin{eqnarray}  \displaystyle{
R P_- > R_s M 
}  \end{eqnarray}
But the product $R P_-$ is the integer $N$, so we find 
\begin{eqnarray} \label{5.5} \displaystyle{
N_{min} \sim R_s M
}  \end{eqnarray}
Another view of the same condition is given in \cite{29}, \cite{9}. 

\subsection{Properties of Schwartzschild black holes} 
Let us now consider the properties of black holes 
that we wish to reproduce. All formulas will be up to 
numerical constants. 

If M-theory is compactified on a $d$-torus it becomes a $D=
11-d$ dimensional theory with Newton constant 
\begin{eqnarray}  \displaystyle{
G_D= \frac{G_{11}}{L^d} = \frac{l_{11}^9}{L^d}
}  \end{eqnarray}
A Schwartzschild black hole of mass $M$ has a radius 
\begin{eqnarray}  \displaystyle{
R_s \sim M^{(\frac{1}{D-3})}G_D^{(\frac{1}{D-3})}
}  \end{eqnarray}
According to Bekenstein and Hawking the entropy of such a 
black hole is 
\begin{eqnarray}  \displaystyle{
S= \frac{\mathrm Area}{4G_D}
}  \end{eqnarray}
where Area refers to the $D-2$ dimensional hypervolume of 
the horizon: 
\begin{eqnarray}  \displaystyle{
{\mathrm Area} \sim R_s^{D-2}
}  \end{eqnarray}
Thus 
\begin{eqnarray}  \displaystyle{
S \sim \frac{1}{G_D} (M G_D)^{\frac{D-2}{D-3}} \sim 
M^{\frac{D-2}{D-3}} G_D^{\frac{1}{D-3}}
}  \end{eqnarray}
Now consider the value of $N_{min} (S)$ given in eq.~(\ref{5.5}): 
\begin{eqnarray}  \displaystyle{
N_{min}(S) = M R_s = M (M G_D)^{\frac{1}{D-3}} = S 
}  \end{eqnarray}
We see that the value of $N_{min}$ in every dimension is 
proportional to the entropy of the black hole. 

In what follows, we will see that the thermodynamic properties of super Yang Mills theory can be estimated by standard arguments 
only if $S {\ \lower-1.2pt\vbox{\hbox{\rlap{$<$}\lower5pt\vbox{\hbox{$\sim$}}}}\ }N$. Thus we are caught 
between conflicting requirements. For $N \gg S$ we don't have 
tools to compute. For $N \ll S$ the black hole will not fit into 
the compact geomrtry. Therefore we are forced to study the 
black hole using $N=N_{min}=S$. 

\subsection{Super Yang Mills thermodynamics}
As we have seen in lecture 2, matrix theory compactified 
on a $d$-torus is described by $d+1$ super Yang Mills theory 
with 16 real supercharges. For $d=3$ we are dealing with a very 
well known and special quantum field theory. In the standard 
3+1 dimensional terminology it is $U(N)$ Yang Mills theory with 4 
supersymmetries and with all fields in the adjoint repersentation. 
This theory is very special. In addition to having 
electric/magnetic duality, it enjoys another property which makes 
it especially easy to analyze, namely it is exactly scale invariant. 

Let us begin by considering it in the thermodynamic limit. The 
theory is characterized by a ``moduli'' space defined by the 
expectation values of the scalar fields $\phi$. Since the $\phi$ 
also represents the positions of the original D0-branes in the 
non compact directions, we choose them at the origin. This 
represents the fact that we are considering a single compact 
object -the black hole- and not several disconnected pieces. 

Now consider the equation of state of the system, defined by 
giving the entropy $S$ as a function of temperature. 
Since entropy is extensive, it is proportional to the volume 
$\Sigma^3$ of the dual torus. Furthermore, the scale invariance 
insures that $S$ has the form 
\begin{eqnarray}  \displaystyle{
S = {\mathrm constant} \: \:  T^3 \Sigma^3 
}  \end{eqnarray}
The constant appearing in this equation counts the number of 
degrees of freedom. Here is what we know about it. For 
vanishing coupling constant, the theory is described by free 
quanta in the adjoint of $U(N)$. This means that the number of 
degrees of freedom is $\sim N^2$. Thus for small $g_{YM}$ 
\begin{eqnarray} \label{5.13} \displaystyle{
S = N^2 T^3 \Sigma^3
}  \end{eqnarray}
Furthermore, for very large $g_{YM}$, the strong/weak 
duality again requires the same equation of state. Although it 
has not been proved, we will assume that eq.~(\ref{5.13}) is 
roughly correct for all $g_{YM}$. 

Finally we may use the standard thermodynamic relation 
\begin{eqnarray}  \displaystyle{
dE=TdS
}  \end{eqnarray}
to obtain the energy of the system:  
\begin{eqnarray}  \label{5.14} \displaystyle{
E \sim N^2 T^4 \Sigma^3 
}  \end{eqnarray}

We will be interested in relating the entropy and mass of the 
black hole. Thus let us eliminate the temperature from 
eq.~(\ref{5.13}) and (\ref{5.14}). We find 
\begin{eqnarray} \label{5.15} \displaystyle{
S = N^2 \Sigma^3 \left( \frac{E}{N^2 \Sigma^3}
\right)^{3/4}
}  \end{eqnarray}

\subsection{Black hole thermodynamics}
Now the energy of the quantum field theory is identified with 
the light cone energy of the system of D0-branes forming the 
black hole. That is 
\begin{eqnarray} \label{5.16} \displaystyle{
E= \frac{M^2}{2 P_-} \approx \frac{M^2}{N} R
}  \end{eqnarray}
Plugging (\ref{5.16}) into (\ref{5.15}) gives 
\begin{eqnarray}  \displaystyle{
S = N^2 \Sigma^3 \left( \frac{M^2 R}{N^3 \Sigma^3}
\right)^{3/4} 
}  \end{eqnarray}
Using eq.~(\ref{3.34}) we obtain 
\begin{eqnarray} \label{5.18} \displaystyle{
S = \frac{1}{N^{1/4}} M^{3/2} \left( \frac{l_{11}^9}{L^3} 
\right)^{1/4}
}  \end{eqnarray}
As we shall see, this formula only makes sense when $N 
{\ \lower-1.2pt\vbox{\hbox{\rlap{$<$}\lower5pt\vbox{\hbox{$\sim$}}}}\ } S$. 

For $N \gg S$ we need more powerful methods to compute 
the equation of state. But, as we have seen, $N \sim S$ is the 
minimal acceptable value for making reliable estimates of 
black hole properties. Thus we evaluate eq.~(\ref{5.18}) 
at $N=S$ to obtain 
\begin{eqnarray}  \label{5.19} \displaystyle{
S = M^{6/5} G_8^{1/5}
}  \end{eqnarray}
This is precisely the correct form for the black hole entropy 
in terms of the mass. 
In order to appreciate the significance of 
this formula let us consider the most general behaviour of $S$ 
consistent with dimensional analysis. 
Since $S$ is dimensionless and $G_8$ has dimensions 
(length$)^6$ we have 
\begin{eqnarray} \label{5.20} \displaystyle{
S= M^a G^b l_{11}^{\: (a-6b)}
}  \end{eqnarray}
There are two undetermined exponents in eq.~(\ref{5.20}). 
Matrix theory gets them both correct! 

The next thing we would like to understand is the temperature of 
the black hole. The standard Hawking temperature is given by 
\begin{eqnarray}  \displaystyle{
T_H = \frac{1}{R_s}
}  \end{eqnarray}
This is the temperature in the rest frame. In the light cone frame 
the temperature is red shifted by a boost factor 
$\displaystyle{\frac{M}{P_-} = \frac{RM}{N}}$. Thus the light 
cone temperature is 
\begin{eqnarray} \label{5.22} \displaystyle{
T_{l.c.} = \frac{R}{R_s} \frac{M}{N}
}  \end{eqnarray}
On the other hand, from (\ref{5.13}) we see that at $N=S$ the 
Yang Mills temperature is 
\begin{eqnarray} \label{5.23} \displaystyle{
T = \left( \frac{1}{N} \right)^{1/3} \frac{1}{\sigma} = 
\frac{RL}{l_{11}^3} \frac{1}{N^{1/3}} 
}  \end{eqnarray}

By identifying $N$ with $S$ and using the $\displaystyle{
R_s^{D-3} = GM}$ relation we find that the temperatures in 
(\ref{5.22}) and (\ref{5.23}) agree.  In other words, the Yang 
Mills temperature $T$ is the Hawking temperature boosted to 
the light cone frame. 

One may wonder why we can not simply use (\ref{5.13}) and 
(\ref{5.14}) for $N > \! \! \! > \! \! \! > S$. If we did so, then 
eq.~(\ref{5.18}) would indicate that black hole entropy would 
depend on $N$ and not just on the mass and Newton constant. 
We can construct an equation of state which does reproduce 
the correct black hole entropy for all $N$. Using eq.~(\ref{5.19}) 
and (\ref{5.16}) we obtain 
\begin{eqnarray}  \displaystyle{
S = E^{3/5} \left( \frac{N}{R} \right)^{5/3} G_8^{1/5}
}  \end{eqnarray}
Using $dE= T dS$ we compute the temperature 
\begin{eqnarray}  \displaystyle{
T= S^{2/3} \frac{R}{N} (G_8)^{-1/3} = S^{2/3} \frac{R}{N} 
\frac{L}{l_{11}^3} = \frac{S^{2/3}}{N} \frac{1}{\Sigma}
}  \end{eqnarray}
or 
\begin{eqnarray} \label{5.25} \displaystyle{
S= (T \Sigma N)^{3/2}
}  \end{eqnarray}

This equation of state agreees with (\ref{5.13}) at $S=N$, but 
also guarantees correct black hole thermodynamics for all 
$N$. The important feature of the equation of state 
(\ref{5.25}) is that the entropy is not extensive. We will return to this shortly. 

A consistent picture can now be formulated. At high 
temperatures for which $S \gg N$ the equation of state is given 
by (\ref{5.13}). The entropy is extensive here because the typical 
wavelenght of a quantum is much smaller than $\Sigma$, the 
size of the dual torus. As the temperature decreases (at fixed 
$N$), eq.~(\ref{5.13}) continues to hold until we come to the 
point 
\begin{eqnarray} \label{5.26a} \displaystyle{
S=N
}  \end{eqnarray}
\begin{eqnarray} \label{5.26b} \displaystyle{
T= \frac{1}{\Sigma N^{1/3}}
}  \end{eqnarray}
At this point the system begins to behave like a black hole 
and, as we have seen, agreement with black hole 
thermodynamics results. However, at this point, a 
transition occurs in the behaviour of $S(T)$. Although we 
don't know how to derive the very low temperature behaviour 
in eq.~(\ref{5.25}), we will see good reason to expect a breakdown of (\ref{5.13})  at the transition point. 

\subsection{Low temperatures} 
Ordinarily, extensivity of the equation of state for a field theory 
will break down when the temperature becomes so low that the 
typical wavelength of a quantum becomes of 
the same order of the size of 
the system. Thus, the equation of state for a free scalar 
field will hold down to temperature 
\begin{eqnarray}  \displaystyle{
T_{min} \approx \frac{1}{\Sigma}
}  \end{eqnarray}
Eq.~(\ref{5.26a}), (\ref{5.26b}) suggest that the bulk equation of state continues to much lower temperature as $N$ becomes 
large. 

Similar behaviour has previously been seen in applying super 
Yang Mills theory to black hole problems 
\cite{Maldacena_Susskind}. 

To illustrate the reason for the equation of state to be continued 
to such low temperature we will consider the example of 1+1 
dimensional super Yang Mills theory. In addition to the usual 
local fields, the degrees of freedom include global 
``Wilson loops'' degrees of freedom. These Wilson loops 
are unitary $N \times N$ matrices which determine how the 
fields transform when they are transported once around 
$\Sigma$. Thus if the Wilson loop satisfies 
\begin{eqnarray}  \displaystyle{
W=1
}  \end{eqnarray}
all fields are periodic. If on the other hand $W$ is not unity, the 
adjoint fields transform as 
\begin{eqnarray}  \displaystyle{
\phi(\sigma + \Sigma) = W^{\dagger} \phi(\sigma) W
}  \end{eqnarray}
Now consider the case where $W$ has the form of a shift matrix 
\begin{eqnarray}  \displaystyle{ 
W= 
\left (\matrix{
0 & 1 & 0 & 0 & \cdots \cr
0 & 0 & 1 & 0 & \cdots \cr
0 & 0 & 0 & 1 & \cdots \cr
0 & 0 & 0 & 0 & \cdots \cr
\cdots & \cdots & \cdots & \cdots & \cdots \cr
1 & 0 & 0 & 0 & \cdots \cr
}\right ) 
}  \end{eqnarray}
The fields satisfy 
\begin{eqnarray}  \displaystyle{
\phi_{a,b} (\sigma + \Sigma) = \phi_{a+1, b+1}(\sigma)
}  \end{eqnarray}

The periodicity of the field $\phi$ is significantly modified. 
In fact the field only returns to its original value after cycling 
around the $\sigma$ axis $N$ times. The result of this is that the 
system can support waves with effective wavelength 
\begin{eqnarray}  \displaystyle{
\lambda = \Sigma N 
}  \end{eqnarray}
and the bulk equation of state continues to temperature of 
order 
\begin{eqnarray}  \displaystyle{
T_{min} \sim \frac{1}{\Sigma N}
}  \end{eqnarray}

In effect, the system behaves as if it lived on a circle of length 
$N$ times larger than its real length. 

In the 3+1 dimensional case of present interest it can be shown 
that Wilson loop configurations exist for which the effective 
volume of the dual torus is $N$ times bigger. In other words the 
dual torus behaves as if it had radii $\displaystyle{ 
\Sigma' = \Sigma N^{1/3}}$. 
Thus it is completely natural that the bulk thermodynamics 
continues down to $\displaystyle{ \Sigma T = \frac{1}{N^{1/3}}}$. 
This represents a remarkable confirmation on the Yang Mills 
side that a transition will happen just when the system fits the 
compact $X^-$ axis. 
\begin{eqnarray}  \displaystyle{
}  \end{eqnarray}
 
\section*{Conclusions} 
In these lectures we have seen the remarkable ways in which the various objects of M-theory and string theory arise out of the 
underlying degrees of freedom of matrix theory. The list 
includes supergravitons, membranes, 5-branes, strings, D-branes and black holes. 

Given this ability to arrange themselves into exactly the correct objects, it seems very likely that matrix theory correctly 
captures the nonperturbative physics of string theory. There are 
however many unresolved questions. First of all, 
compactification on 4, 5 and 6 tori involves non-renormalizable 
field theories. Even worse, the 7-torus introduces major 
difficulties of principle. The problems of probing Lorentz invariance or the existence of the large $N$ limit are unsolved. 
Perhaps most important is that we have no clear general understanding of the connection of Matrix theory and classical 
general relativity. Circumstancial evidence exists that the low energy theory involves Einstein gravity but no clear and 
comprehensive derivation exists at present. Hopefully the 
situation will soon change. 

\section*{Acknowledgements}
This work was supported in part by NSF grant No.~PHY-9219345. 
The authors would like to thank the Institute for Advanced Study 
for hospitality and support during preparation of the article.

\section*{Figures}

\begin{figure}[h]
\epsfxsize=75mm
\epsfbox{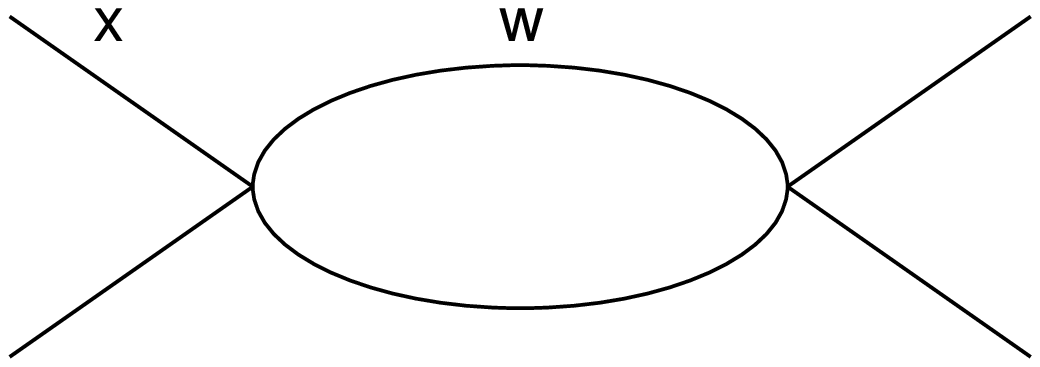}
\end{figure} 
{\large Figure 1.} \\ 

\vspace*{3cm} 
\noindent {\large  Figure 2. (see following page)} 
\begin{figure}[h]
\epsfxsize=75mm
\epsfbox{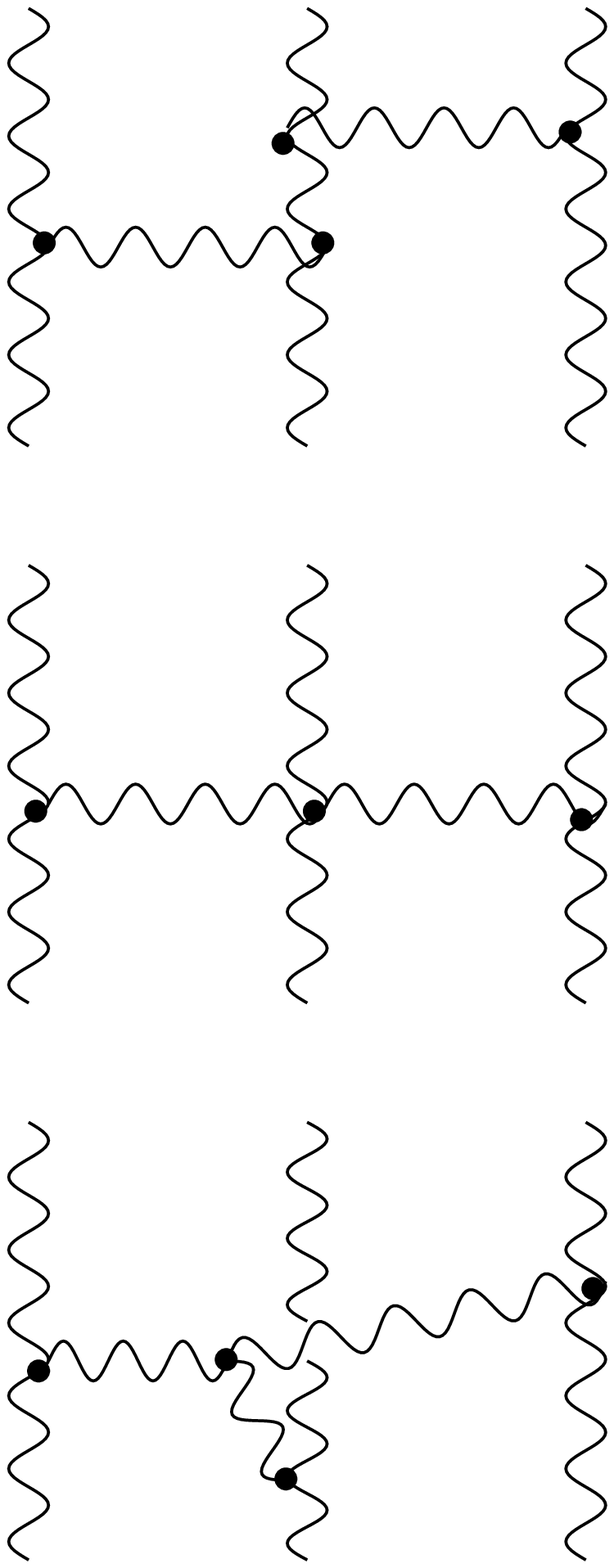}
\end{figure}



\begin{thebibliography}{~~}  
\bibitem{1}{M Theory As A Matrix Model: A Conjecture; 
T.~Banks, W.~Fischler, S.~H.~Shenker, L.~Susskind, 
Phys.~Rev.~{/bf D55} (1997) 5112-5128}
\bibitem{2}{
Green Schwarz and Witten, Superstring theory, 
Cambridge University Press 1987}
\bibitem{3}{The Parton Picture Of Elementary Particles, 
Kogut and Susskind, Phys.~Rept.~{\bf 8} (1973) 75}
\bibitem{4}{On The Quantum Structure Of A Black Hole, 
G.~'t Hooft, Nucl.Phys.~{\bf B256} (1985) 727

Black hole evaporation without information loss, 
C.~R.~Stephens, G.~'t Hooft and B.~F.~Whiting, 
Class.~Quant.~Grav.~{\bf 11} (1994) 621

hep-th/9306069, The stretched horizon and black hole complementarity, L.~Susskind, L.~Thorlacius, and J.~Uglum, 
Phys.~Rev.~{\bf D48} (1993) 3743-3761 } 
\bibitem{5}{hep-th/9506126, Eleven Dimensional Origin of 
String/String Duality: A One Loop Test, 
M.~J.~Duff, J.~T.~Liu, R.~Minasian, 
Nucl.Phys.~{\bf B452} (1995) 261

hep-th/9410167, Unity of Superstring Dualities, 
C.~M.~Hull, P.~K.~Townsend, Nucl.Phys.~{\bf B438} (1995) 109

hep-th/9503124, String Theory Dynamics In Various Dimensions, 
E.~Witten, Nucl.Phys.~{\bf B443} (1995) 85}
\bibitem{6}{hep-th/9508154, Some Relationships Between Dualities in String Theory, Paul S.~Aspinwall, Nucl.~Phys.~Proc.~Suppl.~{\bf 46} (1996) 30-38

hep-th/9510086, The Power of M Theory, John H.~Schwarz, 
Phys.Lett.~{\bf B367} (1996) 97 }
\bibitem{7}{
 hep-th/9409089, The World as a Hologram, L.~Susskind, 
J.~Math.~Phys.~{\bf 36} (1995) 6377 }
\bibitem{8}{hep-th/9704080, Another Conjecture about M(atrix) Theory, 
Leonard Susskind} 
\bibitem{9}{hep-th/9711063, A note on discrete light cone quantization, 
Daniela Bigatti, Leonard Susskind, 
submitted for publication in Physics Letters B} 
\bibitem{10}{hep-th/9710009, Why is the Matrix Model Correct?, 
Nathan Seiberg, Phys.~Rev.~Lett.~{\bf 79} (1997) 3577-3580

hep-th/9709220, D0 Branes on $T^n$ and Matrix Theory, Ashoke Sen }
\bibitem{11}{hep-th/9711037, Compactification in the Lightlike Limit, 
Simeon Hellerman, Joseph Polchinski} 
\bibitem{12}{hep-th/9510017, Joseph Polchinski, 
Phys.~Rev.~Lett.~{\bf 75} (1995) 4724

hep-th/9602052, Notes on D-Branes, Joseph Polchinski, Shyamoli Chaudhuri, 
Clifford V. Johnson} 
\bibitem{13}{
hep-th/9603081, D-particle Dynamics and Bound States, Ulf H. Danielsson, 
Gabriele Ferretti, Bo Sundborg, 
Int.~J.~Mod.~Phys.~{\bf A11} (1996) 5463-5478

hep-th/9603127, A Comment on Zero-brane Quantum Mechanics, 
Daniel Kabat, Philippe Pouliot, Phys.~Rev.~Lett.~{\bf 77} (1996) 1004-1007}
\bibitem{14}{hep-th/9511043, D-brane dynamics, C.~Bachas, 
Phys.~Lett.~{\bf B374} (1996) 37-42}
\bibitem{Polchinski_Pouliot}{hep-th/9704029,  Membrane Scattering with 
M-Momentum Transfer, Joseph Polchinski, Philippe Pouliot, 
Phys.~Rev.~{\bf D56} (1997) 6601-6606}
\bibitem{16}{hep-th/9710174, Multigraviton Scattering in the Matrix Model, 
M.~Dine, A.~Rajaraman} 
\bibitem{17}{hep-th/9705091, A Two-Loop Test of M(atrix) Theory, 
Katrin Becker, Melanie Becker, to appear in Nuclear Physics B} 
\bibitem{Connes}{Alain Connes, Non commutative geometry, 
Academic Press 1994}
\bibitem{18}{hep-th/9610236, Five-branes in M(atrix) Theory, 
Micha Berkooz, Michael R.~Douglas, Phys.~Lett.~{\bf B395} (1997) 196-202} 
\bibitem{19}{hep-th/9612157, Branes from Matrices, 
Tom Banks, Nathan Seiberg, Stephen Shenker, Nucl.~Phys.~{\bf B490} (1997)
91-106 } 
\bibitem{20}{hep-th/9701025, Proposals on nonperturbative superstring 
interactions, Lubos Motl

hep-th/9702187, Strings from Matrices, Tom Banks, Nathan Seiberg, 
Nucl.~Phys.~{\bf B497} (1997) 41-55 

hep-th/9703030, Matrix String Theory, R.~Dijkgraaf, E.~Verlinde, H.~Verlinde, 
Nucl.~Phys.~{\bf B500} (1997) 43-61 }
\bibitem{21}{B.~de Wit, J.~Hoppe and H.~Nicolai, Nucl.~Phys.~{\bf B 305} (1988) }
\bibitem{22}{hep-th/9611042, D-brane field theory on compact spaces, 
Washington Taylor, Phys.~Lett.~{\bf B394} (1997) 283-287}
\bibitem{23}{hep-th/9611164, T Duality in M(atrix) Theory and S Duality in 
Field Theory, Leonard Susskind}
\bibitem{24}{hep-th/9611202, Branes, Fluxes and Duality in M(atrix)-Theory, 
Ori J.~Ganor, Sanjaye Ramgoolam, Washington Taylor IV, 
Nucl.~Phys.~{\bf B492} (1997) 191-204 }
\bibitem{25}{hep-th/9703102, The Incredible Shrinking Torus, 
W.~Fischler, E.~Halyo, A.~Rajaraman, L.~Susskind, 
Nucl.~Phys.~{\bf B501} (1997) 409-426 }
\bibitem{26}{hep-th/9702101, Rotational Invariance in the M(atrix) Formulation of 
Type IIB Theory, Savdeep Sethi, Leonard Susskind, Phys.~Lett.~{\bf B400 (1997)} 
265-268} 
%
\bibitem{27}{hep-th/9705190, Instantons, Scale Invariance and Lorentz Invariance in Matrix Theory, T.~Banks, W.~Fischler, N.~Seiberg, L.~Susskind, 
Phys.~Lett.~{\bf B408 (1997)} 111-116}
%
\bibitem{Seiberg}{hep-th/9705117, Notes on Theories with 16 
Supercharges, Nathan Seiberg}
%
\bibitem{28}{hep-th/9709091, Schwarzschild Black Holes from 
Matrix Theory, T.~Banks, W.~Fischler, I.~R.~Klebanov, 
L.~Susskind

hep-th/9711005, Schwarzchild Black Holes in Matrix Theory II, T.~Banks (Rutgers U.), W.~Fischler (U.~Texas, Austin),
I.~R.~Klebanov (Princeton U.), L.~Susskind (Stanford U.)
       
hep-th/9709108, Schwarzschild Black Holes in Various Dimensions from Matrix Theory, Igor R.~Klebanov, Leonard Susskind (to appear in Physics Letters B)} 
%
\bibitem{29}{hep-th/9710217, Comments on Black Holes in Matrix 
Theory, Gary T.~Horowitz, Emil J.~Martinec} 
%
\bibitem{BFSS2}{hep-th/9705190,Instantons, Scale 
Invariance and Lorentz Invariance in Matrix Theory, 
T.~Banks, W.~Fischler, N.~Seiberg, L.~Susskind, 
Phys.~Lett.~{\bf B408} (1997) 111-116}
%
\bibitem{30}{hep-th/9709114, Black Holes in Matrix Theory, 
M.~Li, E.~Martinec}
%
\bibitem{Maldacena_Susskind}{hep-th/9604042, D-branes and 
Fat Black Holes, Juan M.~Maldacena, 
Leonard Susskind, Nucl.~Phys.~{\bf B475} (1996) 679-690}
%


\end{thebibliography}
\end{document}